\documentclass[aps,pra,twocolumn,showpacs,superscriptaddress]{revtex4-1}
\usepackage{color}
\usepackage{graphicx}
\usepackage{bm}
\usepackage{amsmath}

\begin{document}

\title{Carrier field shock formation of long wavelength femtosecond pulses in dispersive media}

\author{
Paris Panagiotopoulos$^{1,2}$,
Patrick Whalen$^{1,3}$,
Miroslav Kolesik$^{1,2,4}$, 
and Jerome V. Moloney$^{1,2,3}$
\\  
$^1$~Arizona Center for Mathematical Sciences, University of Arizona, Tucson 85721-0094\\
$^2$~College of Optical Sciences, University of Arizona, Tucson 85721-0094 \\
$^3$~Department of Mathematics, University of Arizona, Tucson 85721-0094 \\
$^4$~Constantine the Philosopher University, Nitra, Slovakia}

\date{\today}

\begin{abstract}
We numerically demonstrate the formation of carrier field shocks in various dispersive media for a wide variety of input conditions using two different electric field propagation models. In addition, an investigation of the impact of numerous physical effects on carrier wave shock is performed. It is shown that in many cases a field shock is essentially unavoidable and therefore extremely important in the propagation of intense long wavelength pulses in weakly dispersive nonlinear media such as noble gases, air, and single-crystal diamond. The results presented here are expected to have a significant impact in the field of ultrashort nonlinear optics, attosecond pulse generation, and wavepacket synthesis where the use of mid-IR wavelengths is becoming increasingly more important. 
\end{abstract}

\maketitle

\section{Introduction}

Traditionally, the field of ultrafast nonlinear optics uses near IR lasers for the study of the interaction between atoms and molecules with the intense electric field of the pulse, {\it i.e.} 800 nm for Titanium Sapphire systems utilizing the chirped pulse amplification technique \cite{MourouCPA}. Typically the duration of such wavepackets is on the femtosecond time scale containing multiple electric field cycles.  Therefore, the nonlinear propagation dynamics can be successfully modeled using Schr{\"o}dinger type equations, the most well known of which is the nonlinear envelope equation (NEE) \cite{Couairon1,Brabec}. While NEE based models are well suited to describe most effects relevant to the electric field envelope, they fail by definition to describe carrier wave related effects, most notably for the present work, carrier wave steepening. 

Carrier wave self-steepening and field shock formation have been shown to exist in an ideal dispersionless media through numerical integration of Maxwell's equations \cite{Shock1}, and have very recently been predicted in realistic dispersive media at mid and far-IR wavelengths \cite{Whalen2014} using the modified Kadomtsev-Petviashvili equation (MKP) \cite{Kuznetsov1983,Turitsyn1985,Klein2012,Kozlov1997}. The key factor for this latest result was the use of longer wavelengths that are beneficial both in terms of a higher ionization threshold and an almost flat dispersion landscape. The resulting propagation regime is strongly nonlinear and weakly dispersive, which proves to be ideal for the observation of extreme carrier shocks. On the other hand, the high dispersion and weak nonlinearity found in most transparent media at shorter wavelengths prevents carrier wave steepening from manifesting.

As was shown in Ref.\cite{Whalen2014}, the canonical description for pulse propagation in the strong nonlinear - weakly dispersive regime is the carrier wave resolved modified Kadomtsev-Petviashvili (MKP) equation. MKP models propagate an electric field rather than a wavepacket envelope, hence they are related to the unidirectional pulse propagation equations (UPPE) \cite{Kolesik2004} which are widely used in computational nonlinear optics as an alternative to envelope based models \cite{Practitioners}. In addition, the MKP model is able to predict a variety of propagation related effects that reshape the electric field in spectacular ways, producing exotic "shark-fin" and "top-hat" waveforms. 

The realization that mid and far-IR pulse propagation is very different than near-IR filamentation has far reaching implications since research in the ultrashort optics community has trended towards longer wavelengths in recent years. In Ref.~\cite{Hauri2007}, 2-$\mu$m pulses were used to produce a self-compressed carrier-envelope phase stable few-cycle pulse through filamentation. Furthermore, a recent work by Berg{\'e} \textit{et.~al.}~\cite{Berge2013} illustrated that self-compression efficiency is improved at longer wavelengths; in Berg{\'e}'s study, carrier wavelengths of up to 8-$\mu$m were simulated in argon. Extreme spectral broadening that is observed in the mid-IR regime has been shown to be able to produce keV x-rays \cite{keVxrays} and opens the way for attosecond \cite{attorev} and even zeptosecond pulse generation \cite{zeptoPRL}, and wave-form synthesis \cite{Science2011}.

Given the findings for mid and far-IR wavelength pulses in \cite{Whalen2014,Berge2013}, it is evident that a reassessment of a wide variety of models in multiple disciplines in the field of extreme nonlinear optics is necessary. Full electric field propagators are mandatory for long wavelength, high nonlinearity - low dispersion pulse propagation since carrier field shocks are likely to occur in various media \cite{Whalen2014}. It is clear that propagation effects have to be taken into account when conducting experiments in this regime since the field profile can no longer be considered stationary. Currently, however, the theoretical understanding of the strong nonlinear - weakly dispersive regime is limited. Future studies and analysis of this area will be important as more mid and far-IR mJ femtosecond lasers become commercially available. 

In this work we present an extensive numerical study of the strongly nonlinear - weakly dispersive propagation regime found for long wavelength (mid-IR and far-IR) femtosecond pulses for a variety of materials with two different computational models, the unidirectional pulse propagation equation and the modified Kadomtsev-Petviashvili equation. In our numerical experiments we use a variety of input pulses and materials, while we also isolate various physical effects in order to study their impact on carrier wave shock formation and sustenance. Finally, a comparison of the two theoretical models is used to verify our results. 

\section{Numerical models}

In computational nonlinear optics, solving the full spatio-temporal Maxwell model is known to be extremely expensive in computation time and hence is impractical. Therefore, numerical studies of ultrashort pulse propagation are traditionally conducted either by using envelope type models like the nonlinear Schr{\"o}dinger equation \cite{Couairon1}, or full electric field propagators such as the unidirectional pulse propagation equation (UPPE) \cite{Kolesik2004}. In section~\ref{sec:uppe} we discuss the most general unidirectional field propagation method, the UPPE system. In section~\ref{sec:mkp}, we derive the generalized modified Kadomtsev-Petviashvili (GMKP) equation from the UPPE model, show Keldsyh ionization rates for various media at 8-$\mu$m, and analyze the relevant length scales of the MKP equation. Throughout this letter, results from the GMKP and UPPE models will be presented and compared.

\subsection{Fully spectral solvers: UPPE}
\label{sec:uppe}

In fully spectral solvers, the native representation of the optical field is stored in arrays representing spectral amplitudes that are functions of angular frequency $\omega$, and transverse wavenumbers $k_x,k_y$. For example $\vec{E}_{k_x,k_y,+}(\omega,z)$ would stand for the spectral amplitude in the plane-wave expansion of the field at a given propagation distance $z$.

The spectral nature of the solver makes it possible to solve the linear propagation problem exactly. For a given medium permittivity $\epsilon(\omega)$, the linear propagation properties are encoded in the propagation constants of the modal fields. In the case of a homogeneous medium these are the well-known plane waves, and the propagation  constant depends on both the angular frequency and on the transverse wavenumbers:
\begin{equation}
k_z(\omega,k_x,k_y) \equiv \sqrt{\omega^2 \epsilon(\omega)/c^2 - k_x^2 - k_y^2} \ .
\end{equation}
The permittivity is in general complex-valued, so that not only the chromatic dispersive properties, but also absorption can be accounted for exactly.

Medium properties other than linear  enter the propagation model through the constitutive relations in Maxwell equations. Thus, polarization $\vec{P}(x,y,z,t)$ and the current density $\vec{J}(x,y,z,t)$ must be given as functionals of the electric field  $\vec{E}(x,y,z,t)$. For the sake of brevity, we will not show the dependence of either $\vec{P}$ or $\vec{J}$ on the electric field, because from the point of view of the propagation solver, the nature of this dependence is irrelevant. Let us just note one important aspect and that is that the medium response is calculated from the history of the electric field  $\vec{E}(x,y,z,t)$ independently at a each spatial point. 

The following is an exact system of equations that describes evolution of modal amplitudes along the $z$-axis for the forward and backward propagating field. It is a coupled pair of unidirectional pulse propagation equations (UPPE)~\cite{Kolesik2004}:
\begin{align}
\partial_z \vec{E}^{\perp}_{k_x,k_y,+}(\omega,z) =
+i k_z \vec{E}^{\perp}_{k_x,k_y,+}(\omega,z) + \nonumber \\
\sum_{s=1,2} \vec e^{\perp}_s \vec e^{\phantom{\perp}}_s . [ \frac{i\omega^2}{2 \epsilon_0 c^2 k_z} \vec{P}_{k_x,k_y}(\omega,z)  -
 \frac{\omega}{2 \epsilon_0 c^2 k_z}\vec J_{k_x,k_y}(\omega,z)]
\label{eq:z1}
\end{align}
which describes the forward propagating wave, and
\begin{align}
\partial_z \vec{E}^{\perp}_{k_x,k_y,-}(\omega,z) =
-i k_z \vec{E}^{\perp}_{k_x,k_y,+}(\omega,z)
- \nonumber \\
\sum_{s=1,2} \vec e^{\perp}_s \vec e^{\phantom{\perp}}_s . [ \frac{i\omega^2}{2 \epsilon_0 c^2 k_z} \vec{P}_{k_x,k_y}(\omega,z)  -
 \frac{\omega}{2 \epsilon_0 c^2 k_z}\vec J_{k_x,k_y}(\omega,z)]
\label{Xeq:z2}
\end{align}
which represents the backward propagating radiation.

The above UPPEs are mutually coupled through the medium response represented by the polarization and current density terms. As noted above, they require the knowledge of the full (i.e. forward {\em and} backward) electric field. This is where the unidirectional approximation must be adopted. It requires that the medium response can be accurately calculated from only the forward portion of the field, i.e.:
\begin{equation}
\vec P(\vec E_- + \vec E_+), \vec J(\vec E_- + \vec E_+) \to 
\vec P(\vec E_+), \vec J(\vec E_+)
\label{eq:approxz}
\end{equation}
If this is the case, then it is sufficient to solve only the UPPE representing the forward wave. Obviously, in practice {\em all} pulse propagation simulations, independently of the propagation model they utilize, require this approximation.
The physical meaning of ``unidirectionality'' was discussed by Kinsler in \cite{kinsler_theory_2005,kinsler_limits_2007,kinsler_optical_2010,kinsler_unidirectional_2010} and a practical way to quantify and test its limitations is described in~\cite{unilimits}. The canonical form of the scalar UPPE is as follows (for brevity, we consider the current term grouped with the nonlinear polarization and omit the component and direction subscripts):
\begin{align}
\partial_z E_{k_x,k_y}(\omega,z) =
i K&(k_x,k_y,\omega) E_{k_x,k_y}(\omega,z) \nonumber \\
&+ i Q(k_x,k_y,\omega) P_{k_x,k_y}(\omega,z)  
\phantom{XX}
\label{eq:z2}
\end{align}
with the exact linear propagator
\begin{equation}
K(k_x,k_y,\omega) = \sqrt{ \omega^2 \epsilon(\omega)/c^2 - k_x^2 - k_y^2 }
\end{equation}
and the nonlinear response coupling
\begin{equation}
Q(k_x,k_y,\omega) \equiv \frac{\omega^2}{2 \epsilon_0 c^2   \sqrt{ \omega^2 \epsilon(\omega)/c^2 - k_x^2 - k_y^2 }  } \ .
\end{equation}
This is in fact the form most often utilized in practical simulations.

\subsection{MKP}
\label{sec:mkp}
While the canonical UPPE predicts carrier field shock for long-wavelength high power lasers, this effect and associated scales are hidden in the general form of the model itself. To reveal the physics in a more explicit way, we make several approximations to equation \eqref{eq:z2} that are valid for a large class of filamentation problems yet reduce the model to a simpler and easier to analyze form. 

First, we apply the following Taylor expansions to $K(k_x,k_y,\omega)$ and $Q(k_x,k_y,\omega)$
\begin{align*}
k_{z}(\omega,k_x,k_y) &= k(\omega) \left(1 - \frac{k_{\perp}^2}{2 k^2(\omega)} - \frac{k_{\perp}^4}{8 k^4(\omega)} - \cdots \right) \\ 
\frac{1}{k_{z}(\omega,k_x,k_y)} &= \frac{1}{k(\omega)} \left(1 + \frac{k_{\perp}^2}{2 k^2(\omega)} + \cdots \right),
\end{align*}
with $k_{\perp}^2 = k_x^2 + k_y^2$, and $k(\omega)=\omega \sqrt{\epsilon(\omega)}/c$. Then our expanded UPPE is
\begin{multline} 
\label{eq:tayUPPE} 
  \partial_{z} E_{k_x,k_y}(\omega,z)  = 
   ik(\omega)  E_{k_x,k_y}(\omega,z) \\ 
  -  \frac{ick^{2}_{\perp}}{2n(\omega)\omega} E_{k_x,k_y}(\omega,z) 
   +   \frac{i \omega} {2  \epsilon_{0} c  n(\omega)} P_{k_x,k_y}(\omega,z) \\
   - \frac{i c k_{\perp}^2} {8 n(\omega) \omega}\frac{k_{\perp}^2}{k^2(\omega)} E_{k_x,k_y}(\omega,z)  \\
   + \frac{i \omega} {4  \epsilon_{0} c  n(\omega)}\frac{k_{\perp}^2}{k^2(\omega)} P_{k_x,k_y}(\omega,z) 
   + h.o.t.  
\end{multline}
where $h.o.t.$ denotes the higher order terms in the respective expansions of $k_{z}$ and $k_{z}^{-1}$. Restricting ourselves to paraxial beam propagation implies that only the  first three right-hand-side terms must be retained. Furthermore, $n^{-1}(\omega)$ in the remaining terms can be Taylor expanded around a reference frequency $\omega_{R}$ such that 

\begin{align}
  \partial_{z} & E_{k_x,k_y}(\omega,z) =  i k(\omega) E_{k_x,k_y}(\omega,z)  \nonumber \\ 
  &- \frac{ick^{2}_{\perp}}{2n(\omega_R)\omega}\left[1-\frac{n'(\omega_R)(\omega-\omega_R)}{n(\omega_R)}
  +  \cdot \cdot \cdot \right] E_{k_x,k_y}(\omega,z)  \nonumber \\ 
  &+ \frac{i \omega } {2  \epsilon_{0} c
  n(\omega_R)} \left[1-\frac{n'(\omega_R)(\omega-\omega_R)}{n(\omega_R)} + \cdot \cdot \cdot
  \right]  P_{k_x,k_y}(\omega,z) .  
\end{align}
Now we invoke a restriction on the medium and spectral bandwidth of pulses admitted to our model
\[ \left |  n'(\omega_R)(\omega - \omega_R) \right | \ll {n(\omega_R)}. \]
This is akin to the slowly-evolving-wave approximation (SEWA)
\begin{align*}
  \left |  n'(\omega_R)\omega_R \right | \ll {n(\omega_R)},
\end{align*}
which is used in envelope models to accurately describe light propagation down to the single-cycle regime \cite{Brabec}. Our approximation is somewhat more demanding as it puts additional restrictions on the medium when propagating high harmonic generating pulses; as the harmonic order increases, $\omega$ will deviate more from $\omega_R$ and hence $n'(\omega_R)$ will need to be small enough to account for this. 
After approximations, the resulting propagation equation is
\begin{align} 
\label{eq:neeMKP} 
  \partial_{z}  & E_{k_x,k_y}(\omega,z)  =  ik(\omega) E_{k_x,k_y}(\omega,z) \nonumber \\
  & - \frac{ick^{2}_{\perp}}{2n(\omega_R)\omega}  E_{k_x,k_y}(\omega,z)
   + \frac{i \omega }{2 \epsilon_{0} c  n(\omega_R)}P_{k_x,k_y}(\omega,z) .  
\end{align} 

\begin{figure}
\includegraphics[width=8.6 cm]{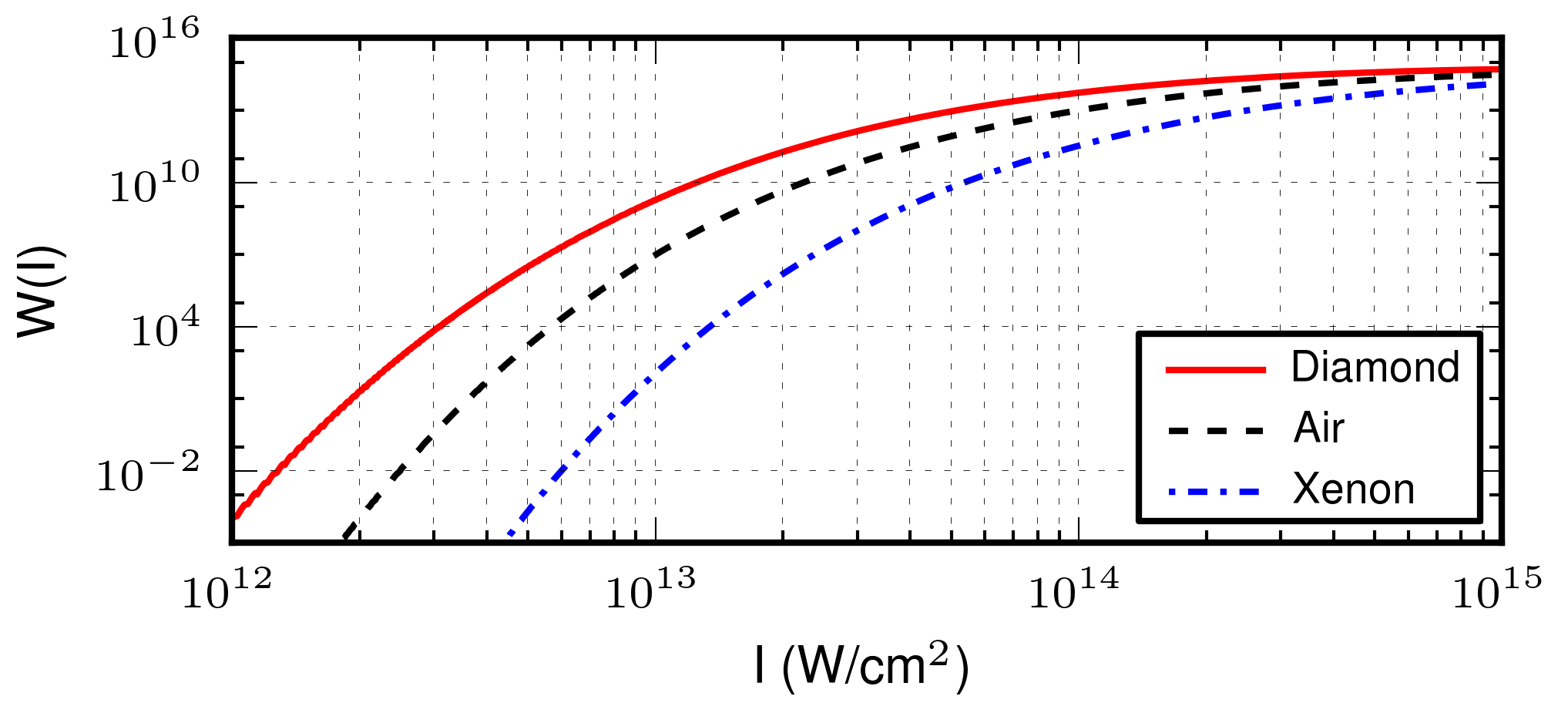}
\caption{(Color online) Keldysh photoionization rates for air (approximated by that of oxygen), diamond, and xenon at a
carrier wavelength of $8 \mu$m.}
\label{Fig:keldysh}
\end{figure}

The following dispersion relation approximation is utilized
\begin{align} 
\label{eq:MKPdispK} 
  k(\omega) = a \omega^{3} - \frac{b}{\omega} + q \omega. 
\end{align}  
While no explicit $\omega_R$ is present in the dispersion relation, equation \eqref{eq:MKPdispK} is a local approximation as the parameters $a$, $b$, and $q$ are chosen to accurately reflect the full dispersion profile (e.g. Sellmeier, etc.) near a specific frequency $\omega_R$. Each example in this work considers a single laser pulse launched with a carrier frequency $\omega_0$ corresponding to a wavelength $\lambda_0$. Therefore, we take the reference frequency as $\omega_R = \omega_0$ and let $n_0 = n(\omega_R)$. Upon substitution into equation \eqref{eq:neeMKP} and transforming to time-domain real-space, the propagation model simplifies to 
\begin{align} 
\label{eq:MKPgen1} 
& \partial_{\tau} \left( \partial_z E  - a \partial_{\tau}^{3}E
    + \frac{1}{2 \epsilon_{0} c n_0}\partial_{\tau} P \right ) \nonumber \\ 
 & \quad \quad \quad  + b E = \frac{c}{2 n_0} \Delta_{\perp} E, 
\end{align} 
where $\tau = t-qz$ and the fields are functions of $(x,y,\tau,z)$. At this point it is easy to separate the nonlinear current density from the nonlinear polarization by replacing $\partial_{\tau} P$ with $\partial_{\tau} P + J$. Furthermore, we split $J$ into a lossless plasma defocusing term and a nonlinear absorption term
\begin{align}
J &= J_{plas} + J_{abs}, \\
\partial_{\tau} J_{plas} &= \frac{e^2}{m_e} \rho E,  \\ 
\frac{J_{abs}}{\epsilon_0 c n_0} &= \frac{W(I)U_i}{I} (\rho_{nt} - \rho) E,  \\
\partial_{\tau}\rho &= W(I)(\rho_{nt} - \rho),
\end{align}
where $I = \epsilon_0 c n_0 <E^2>$ is the intensity of electric field, $U_i$ is the potential, $\rho_{nt}$ is the neutral plasma density, and $W$ is the ionization rate of free charge $\rho$. We also use an instantaneous Kerr nonlinear response such that
\begin{align*}
P =  \epsilon_0 \chi^{(3)} E^{3} = \frac{4}{3} \epsilon_0 (\epsilon_0 c n_0^2 n_2) E^3.
\end{align*}
This produces a propagation model with the electric field in units of $V/m$,
\begin{multline} 
\label{eq:MKPgen2} 
\partial_{\tau} \left( \partial_z E  - a \partial_{\tau}^{3}E
    + \frac{\chi^{(3)}}{2 c n_0}  \partial_{\tau} E^3 
    + \frac{1}{2 \epsilon_{0} c n_0} J_{abs} \right ) \\ 
    + \left(b + \frac{e^2}{2 \epsilon_0 n_0 c m_e} \rho \right) E
          = \frac{c}{2 n_0} \Delta_{\perp} E. 
\end{multline} 
Rescaling $E$ by
\[ \tilde{E} = \left( \frac{\epsilon_0 c n_0}{2} \right)^{1/2} E, \]
and removing tilde's after substitution, we obtain the generalized modified Kadomtsev-Petviashvili (GMKP) optical filamentation equation
\begin{align} 
\label{eq:MKPgen} 
   \partial_{\tau} \left( \partial_z E - a \partial_{\tau}^3 E
   + \frac{4 n_2}{3 c} \partial_{\tau} E^3  + \mathcal{N}_{abs}(\rho,I) E \right) \nonumber \\
            + \left(b + b_{plas}(\rho) \right) E
   = \frac{c}{2 n_0} \Delta_{\perp} E, 
\end{align}
with
\begin{align}
 b_{plas}(\rho) &= \frac{e^2}{2 \epsilon_0 n_0 c m_e} \rho, \\
 \mathcal{N}_{abs}(\rho,I) &= \frac{W(I) U_i  (\rho_{nt} - \rho)}{2 I}.
\end{align}
At long-wavelengths, a full Keldysh ionization response $W(I)$ is preferable to the multiphoton approximation \cite{Berge2013}. In Fig.~\ref{Fig:keldysh}, we show Keldysh photoionization rates for air (approximated by oxygen), diamond, and xenon at a carrier wavelength of $8 \mu$m. These ionization rates are computed from formulas derived in \cite{Mishima2002}. 

The modified Kadomtsev-Petviashvili (MKP) equation is generated by dropping the nonlinear current density terms and ignoring anomalous dispersion ($b = 0$). With these simplifications, we have
\begin{align} 
\label{eq:MKPstand} 
   \partial_{\tau} \left( \partial_z E - a \partial_{\tau}^3 E
   + \frac{4 n_2}{3 c} \partial_{\tau} E^3 \right) 
          = \frac{c}{2 n_0} \Delta_{\perp} E.
\end{align} 
Its instructive to normalize and non-dimensionalize this equation with
\begin{align*} 
  U = \frac{E}{\sqrt{I_{0}}}, && \tilde{\tau} = \omega_0 \tau,&& \tilde{r} = \frac{r}{w_{0}}, &&  \tilde{z} = \frac{z}{L},  
\end{align*} 
where $I_0$ is the initial intensity, $w_0$ is the beam diameter, and $L$ is an arbitrary length scale. After substitutions we obtain
\begin{align} 
\label{eq:ncvecMKP} 
  \partial_{\tilde{\tau}} \left( \partial_{\tilde{z}} U - \frac{L}{12 L_{CSD}} \partial_{\tilde{\tau}}^3 U
   + \frac{4L}{3L_{NL}} \partial_{\tilde{\tau}}  U^3 \right) 
   =\frac{L}{4 L_{DF}} \Delta_{\tilde{\perp}} U, 
\end{align}
with length scales
 \begin{align} 
  \label{eq:lengthScale1} 
    L_{DF} = \frac{k(\omega_0)w_{0}^{2}}{2}, &&  L_{NL} = \frac{c}{\omega_0 n_{2}I_{0}}, 
     && L_{CSD} = \frac{1}{12 a \omega_0^3}.
  \end{align} 
The diffractive length scale $L_{DF}$ and nonlinear length scale $L_{NL}$ are standard length scales in nonlinear optics that describe the propagation of an electric field envelope. On the other hand, the carrier shock dispersion length scale $L_{CSD}$ describes the propagation of the electric field, not the envelope; in particular, $L_{CSD}$ measures the distance over which carrier wave steepening dissipates. With the MKP dispersion relation \eqref{eq:MKPdispK}, $L_{CSD}$ can alternatively be written as \[ L_{CSD} = \frac{2}{k(3\omega_0)-3 k(\omega_0)}, \] which clearly relates the carrier shock dispersion to the coherence between the fundamental and the third harmonic. This important observation connects carrier wave shock to harmonic generation and harmonic walk-off. A complete discussion of $L_{CSD}$ is given in section \ref{sec:fieldShockChar}.

\subsection{UPPE vs GMKP}
The essential difference between the UPPE and GMKP models is in the way dispersion is modeled; the UPPE accurately represents dispersion over a global frequency range whereas the MKP dispersion, as shown in Eq.~\eqref{eq:MKPdispK}, is a local approximation whose accuracy depends on the the carrier frequency. In addition, the UPPE model takes into account vibrational Raman, the full complex $\chi$ relation including transmittance, and plasma losses due to a finite electron collision time. As will be shown later on, even though UPPE is a more complex and complete model for nonlinear wave propagation, the physics that govern carrier shock formation can be adequately described by the simpler GMKP model.

Unless stated otherwise, results presented in this work are obtained using the GMKP model. However, the UPPE is used to compare GMKP results with the complete model (Fig.~\ref{Fig_14}) and to investigate effects either not accounted for in the GMKP model, or those that would push the GMKP beyond its validity limits. In particular, the UPPE is used to study the impact on carrier field shock due to Raman (Fig.~\ref{Fig_11}), high numerical aperture focusing (Fig.~\ref{Fig_6}), and artificially high plasma generation (Fig.~\ref{Fig_13}).

\section{Media and pulse description}

The numerical experiments are conducted in a variety of modeled materials ranging from atomic gases to solid crystal diamond. The common denominator in all cases is the flat dispersion curve at mid and long IR wavelengths (from $\sim$ 4-$\mu$m  up to 8-$\mu$m) that is essential for the formation of a carrier field shock.

In our previous work we presented the first prediction of carrier shock formation in noble gases \cite{Whalen2014}. In this follow-up work, we expand the study of field shock formation to other materials such as air and single-crystal diamond. However, since noble gases are a very attractive option for field shock formation, a complementary study was performed in order to accurately predict the regime in which field shocks are likely to occur (section V). The dispersive and nonlinear properties of the noble gases were taken from \cite{XeDS,Xen2}.

Air is a more complex gas since it consists of oxygen and nitrogen, and exhibits vibrational and rotational nonlinear responses. Nevertheless, it is another good candidate for carrier shock formation due to the flat dispersive curve at longer wavelengths \cite{AirDS}. The nonlinear refractive index is  $n_2=3.2\times 10^{-19}$ cm$^2$/W \cite{Airn2}, and the ionization potential is taken to be that of oxygen $U_i=9.0  eV$ \cite{Couairon1}, which is very close to Xenon. Although air has a Raman response that is well known \cite{Raman1, Raman2}, it will not be taken into account in this work. The effect of Raman will be investigated for the case of single-crystal diamond, as will be discussed below. 

\begin{figure}
\includegraphics[width=8.6 cm]{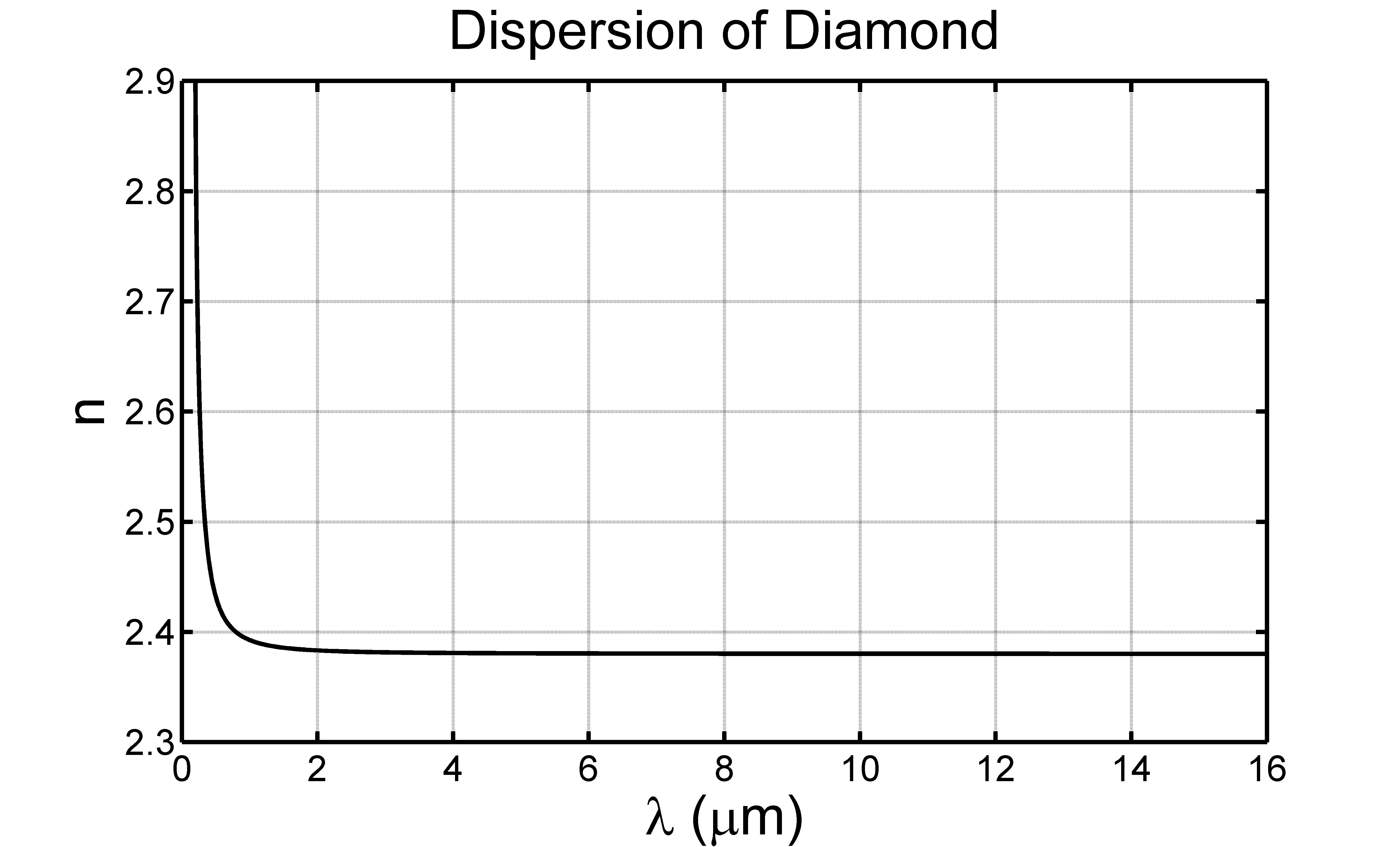}
\caption{Refractive index of single-crystal diamond as a function of wavelength.}
\label{Fig_1}
\end{figure}

\begin{figure}
\includegraphics[width=8.6 cm]{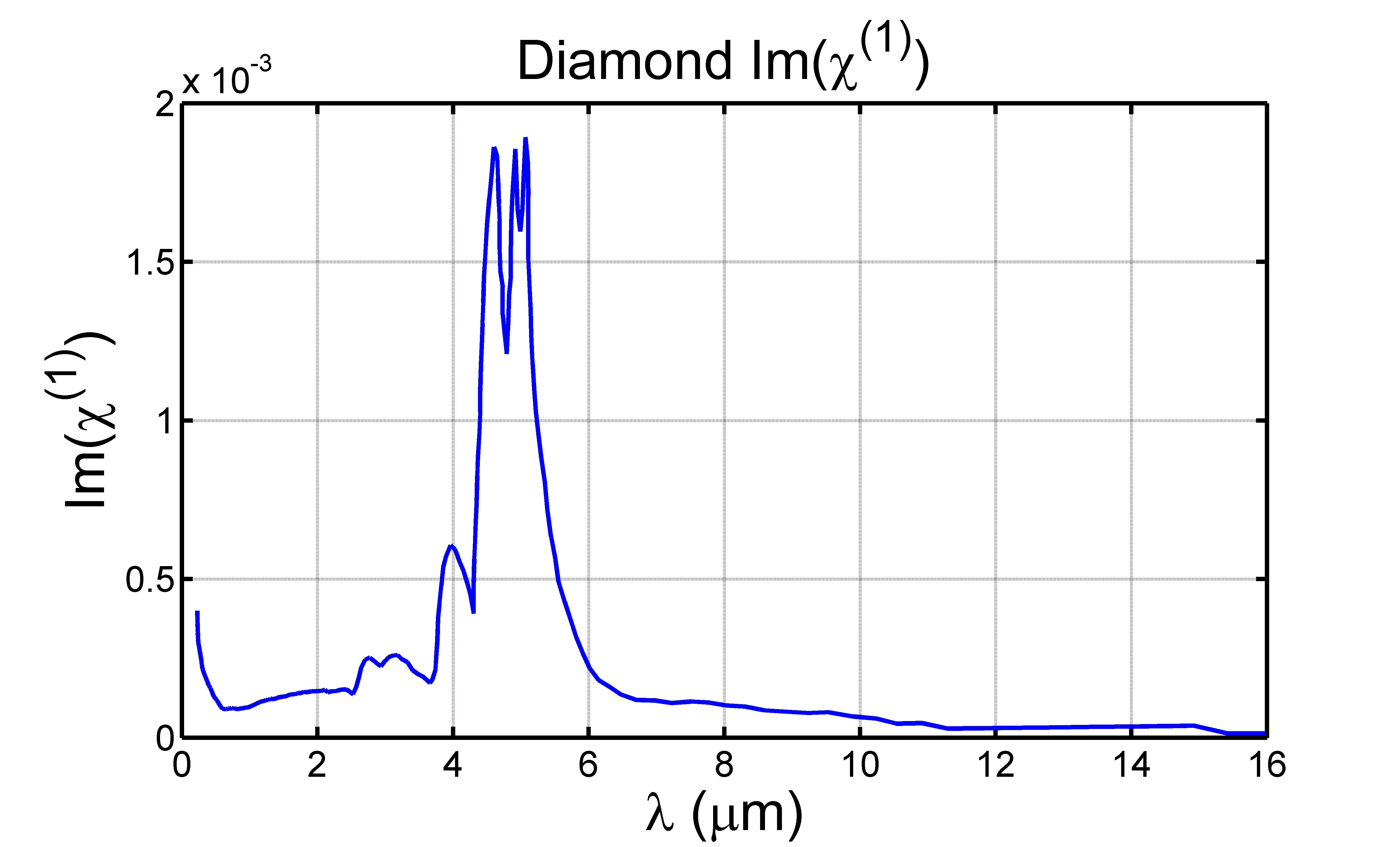}
\caption{(Color online) Imaginary value of $\chi$ of single-crystal diamond as a function of wavelength. Note the two-phonon absorption peaks between 4 $\mu$m and 6 $\mu$m.}
\label{Fig_2}
\end{figure}

In addition to gaseous media, solids can potentially be suitable for producing carrier shocks due to their higher nonlinear responses. Concerning nonlinear field reshaping, diamond is in fact a close crystalline analogy to an atomic gas like Xenon. Single crystal diamond is a material not often used in the field of nonlinear optics, however it is ideal for the purpose of studying carrier field shocks. The main reasons for this are: First, diamond has a relatively high nonlinear refractive index of $n_2=1.3 \times 10^{-15}$ cm$^2$/W \cite{Boyd} (about one order of magnitude higher than fused silica glass), which is important since field shock formation is essentially a nonlinear phenomenon. Second, diamond has an almost flat dispersion curve at longer wavelengths \cite{Zeitsev2001,Edwards81} as can be seen in Fig.~\ref{Fig_1}. In addition, the imaginary part of the susceptibility $\chi$ is almost flat if not for the weak two phonon absorption lines around 5-$\mu$m \cite{Kaminskii2006}, shown on Fig.~\ref{Fig_2}. Third, diamond has a relatively high ionization potential for a solid crystalline material of $U_i=5.5 eV$, which makes the ionization relatively weak at longer wavelengths. Diamond will also be used to assess the effect of vibrational Raman on field shock formation. The Raman frequency $\omega=2.5099 \times 10^{14}$ s$^{-1}$ and time scale $\tau=4.2$ ps are taken from \cite{DiaRaman,DiaRaman2}. Our simulations have included most of the relevant effects needed to model nonlinear propagation of femtosecond pulses in diamond. These simulations support our argument that diamond is an attractive material for electric field steepening (see Fig.\ref{Fig_5}, for example). Furthermore, we expect that results presented here will motivate future experiments utilizing diamond as a material in the study of nonlinear phenomena including but not limited to field shocks.

Other, less exotic solids, like fused silica or BK7 glass, have proven to be unsuitable to produce a field shock. The reason for this is the high and even anomalous dispersion they exhibit in the IR, which will be highlighted later in section V.

Throughout this work, a Gaussian spatio-temporal laser pulse is used, with a wavelength of $\lambda_0 = 8$ $\mu$m and a pulse duration of $\tau_p =$ 40 fs at $1/e^2$ radius. Unless stated otherwise, the starting peak intensity was set to be $I_0=4 \times 10^{12}$ W/cm$^2$, while the beam is collimated with a waist of $w_0 = 2$ mm at $1/e^2$ radius for air, and $w_0 = 50$ $\mu$m for diamond. All simulations, with the exception of those corresponding to Fig.~\ref{Fig_9}, are done in three dimensions (3D): with assumption the of a cylindrically symmetric solution in $r$, time $t$, and propagation distance $z$.

\section{Field shock formation and evolution}
\begin{figure}
\includegraphics[width=8.6 cm]{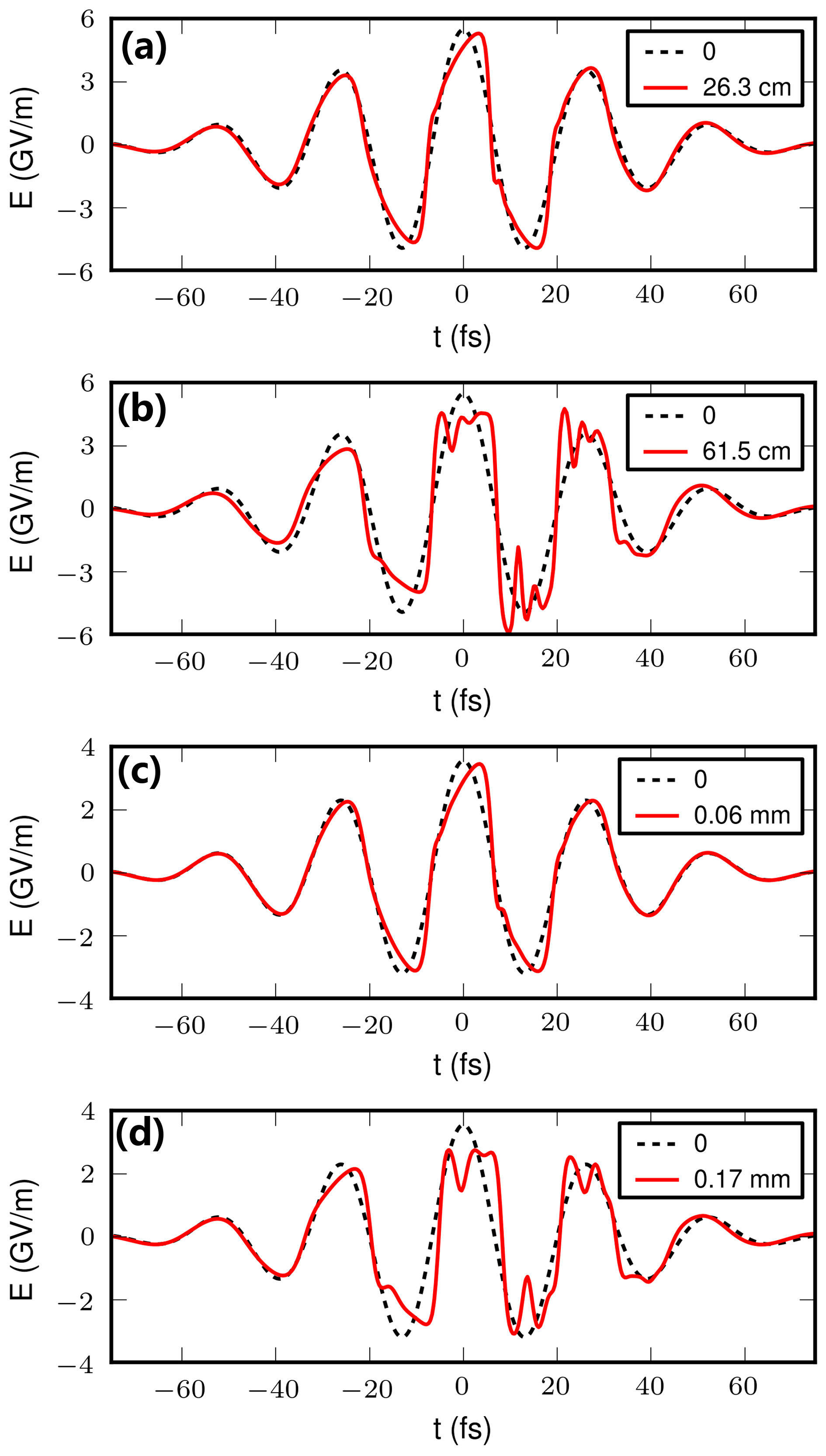}
\caption{(Color online) Electric field of an 8-$\mu$m pulse propagating in air (a and b), and diamond (c and d). Black dashed line: initial field shape, red lines: field shape at (a) $z=26.3$ cm and (b) $z=61.5$ cm, for air, and (c) $z=60$ $\mu$m and (d) $z=170$ $\mu$m, for diamond.}
\label{Fig_3}
\end{figure}

In Fig. \ref{Fig_3} we can see the generated carrier wave shock in air (a, b) and diamond (c, d). The input power (beam waist) and propagation distances are different in each case to compensate for the material properties but the end results are very similar. In both cases (a, c) the electric field of the laser pulse is undergoing extreme steepening and is reshaped into a "shark-fin" like temporal structure. After additional propagation, both fields continue to evolve and are transformed into almost top-hats (b, d). These extreme nonlinear dynamics acting on the carrier wave, in two totally different materials, indicate that the mechanisms leading to carrier wave shock formation are universal in the strongly nonlinear - weakly dispersive long wavelength regime. 

\begin{figure}
\includegraphics[width=8.6 cm]{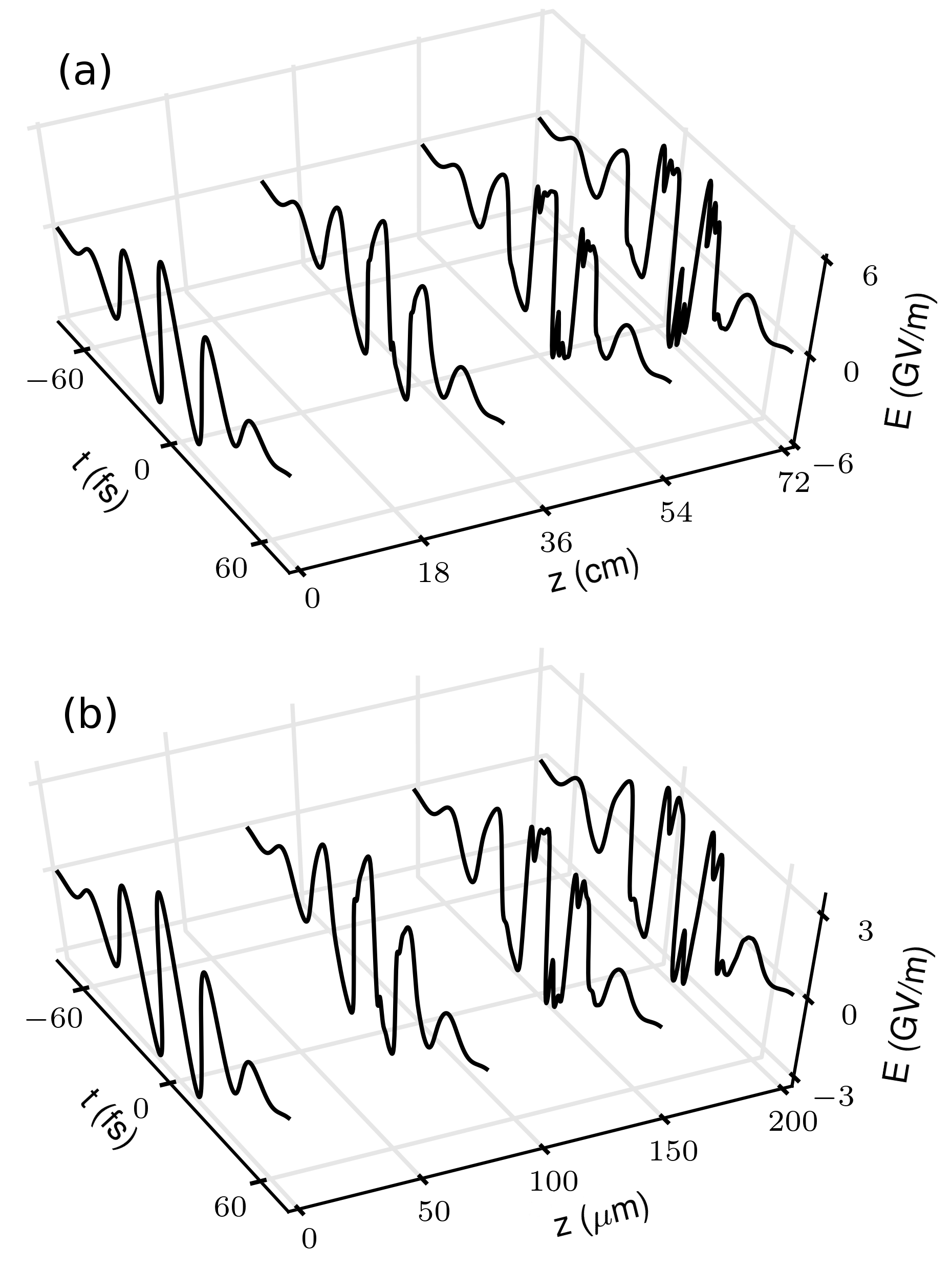}
\caption{On-axis electric field evolution in (a) air, and (b) single-crystal diamond at $\lambda_0= 8$ $\mu$m. In both cases, $I_0 = 4.0 \times 10^{12}$ W/cm$^2$ and $t_p = 40$ fs, with $w_0 = 2$ mm for air and $w_0 = 50$ $\mu$m for diamond. The electric fields initially evolves into a shark-fin shock with harmonic walk-off further reshaping the fields over longer distances.}
\label{Fig_4}
\end{figure}

It is important to stress that in this long wavelength - highly nonlinear - low dispersive regime, propagation effects must be accounted for. As we can see in Fig. \ref{Fig_4}, where the evolution of the electric field of our wavepacket in air and diamond is shown, the field shape is in fact constantly evolving as the pulse is propagating in z. In both cases, after the initial steepening which can be seen in (a) at $z=30$ cm and (b) at $z=80$ $\mu$m, the field is constantly reshaping before finally taking on a top-hat shape at (a) $z=72$ cm, and (b) $z=200$ $\mu$m. Obviously the rate at which the field evolves is directly proportional to the characteristic lengths $L_{CSD}$ and $L_{NL}$. The continuous broadening of the spectrum and walk-off of the numerous harmonics results in a dynamically evolving wave-form, which can change drastically over just a few microns of propagation. Therefore, any experiments and numerical models related to nonlinear field shaping \cite{keVxrays,attorev,zeptoPRL,Science2011} should be reassessed given the results presented here. 

\begin{figure}
\includegraphics[width=8.6 cm]{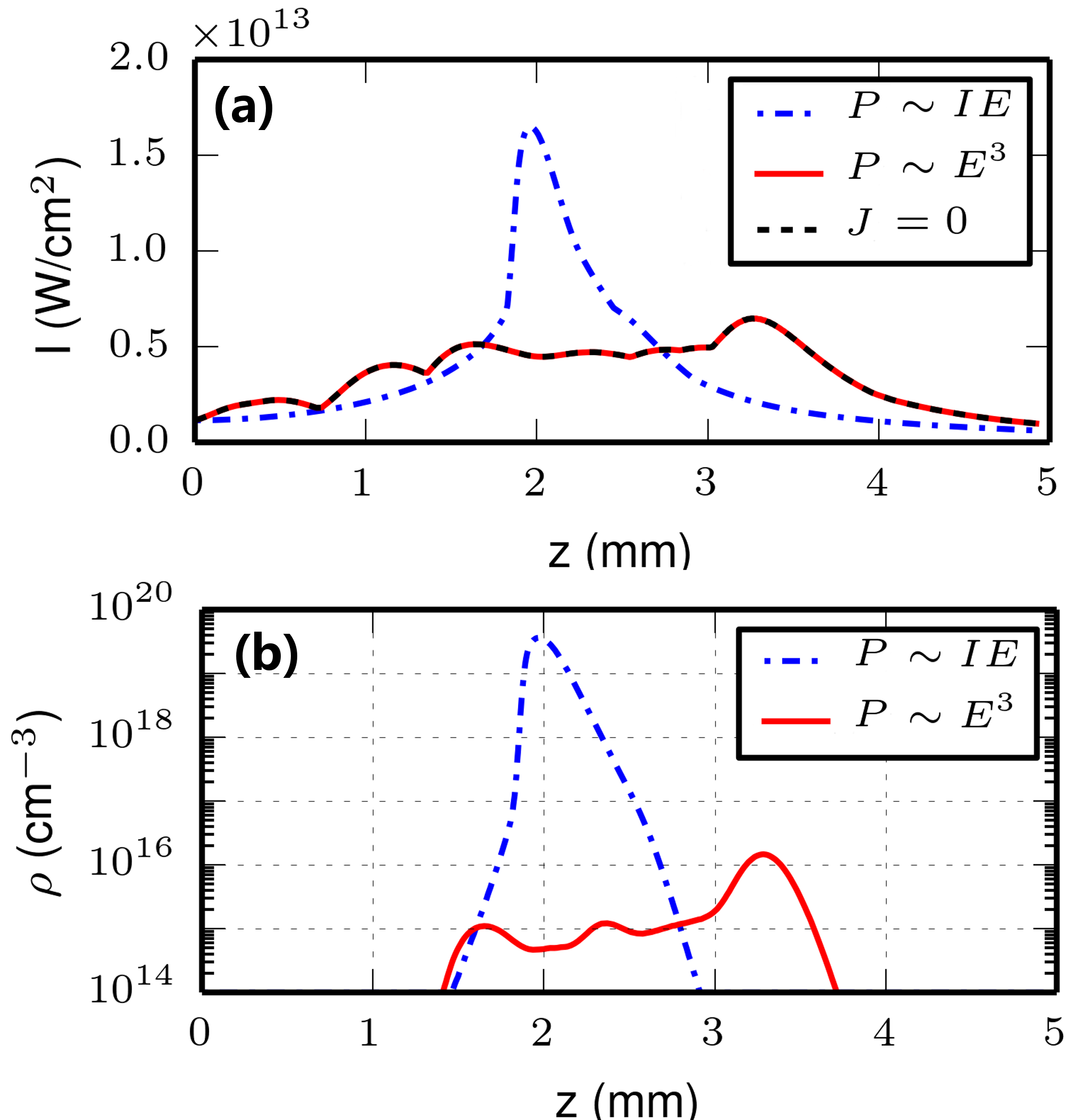}
\caption{(Color online) Critical collapse of a $\lambda_0 = 8$ $\mu$m Gaussian wavepacket propagating in single-crystal diamond with an input power of $P_{in} = 1.46 P_{cr}$. Collapse arrest due to plasma ($P \sim IE$) and carrier shock dispersion ($P \sim E^3$) are illustrated in plots of (a) peak intensity vs. propagation distance, and (b) peak plasma vs. propagation distance.}
\label{Fig_5}
\end{figure}

Another interesting observation in this low ionization regime is that collapse can still be arrested, this time due to the walk-off of the self-generated higher harmonics. This is clearly observed in Fig.~\ref{Fig_5}(a) where the peak intensity along propagation in single-crystal diamond is shown (red continuous line). The extremely sharp field gradients which evolve due to the low dispersion -  high nonlinearity lead to even more spectral broadening. This eventually results in the spreading of a significant portion of the total energy over a large spectral region, effectively dispersing the wavepacket and arresting collapse. Note that the role of dispersion is key; on the one hand, very weak dispersion allows for a carrier wave shock to manifest itself, on the other hand, non-zero dispersion is arresting the collapse and eventually breaking up the field profile itself. The role of plasma in this collapse arrest is actually very small. As observed in Fig.~\ref{Fig_5}(b) (red continuous line), the generated plasma density is essentially kept under $\sim 10^{16}$ cm$^{-3}$, which is rather weak in the filamentation regime. To further test this, we switched plasma generation off entirely ($J=0$) and observed that the peak intensity of the wavepacket was essentially unaffected (Fig.~\ref{Fig_5}(a)). That is, carrier shock dispersion is the primary mechanism by which critical collapse is arrested, in contrast to a shorter wavelength regime in which the main arrest mechanism is due to plasma.

The role of field dispersion and envelope dispersion are significantly different here. When wavepacket propagation is simulated using an envelope type nonlinear Kerr response ($P \sim IE$), instead of a carrier-resolving Kerr response ($P \sim E^3$), the collapse arrest is caused by optical field ionization and plasma defocusing. This can be seen in Fig.~\ref{Fig_5} where the peak intensity (a) and peak electron density (b) are plotted in the blue dash-dotted lines. The results shown in Fig.~\ref{Fig_5} showcase again the importance of the use of field propagators for mid-IR wavelengths; envelope type models are unable to properly describe the evolution of the wavepacket, predicting collapse arrest due optical field ionization rather than carrier shock dispersion, and overestimating plasma generation and nonlinear losses in the process.

\begin{figure}
\includegraphics[width=8.6 cm]{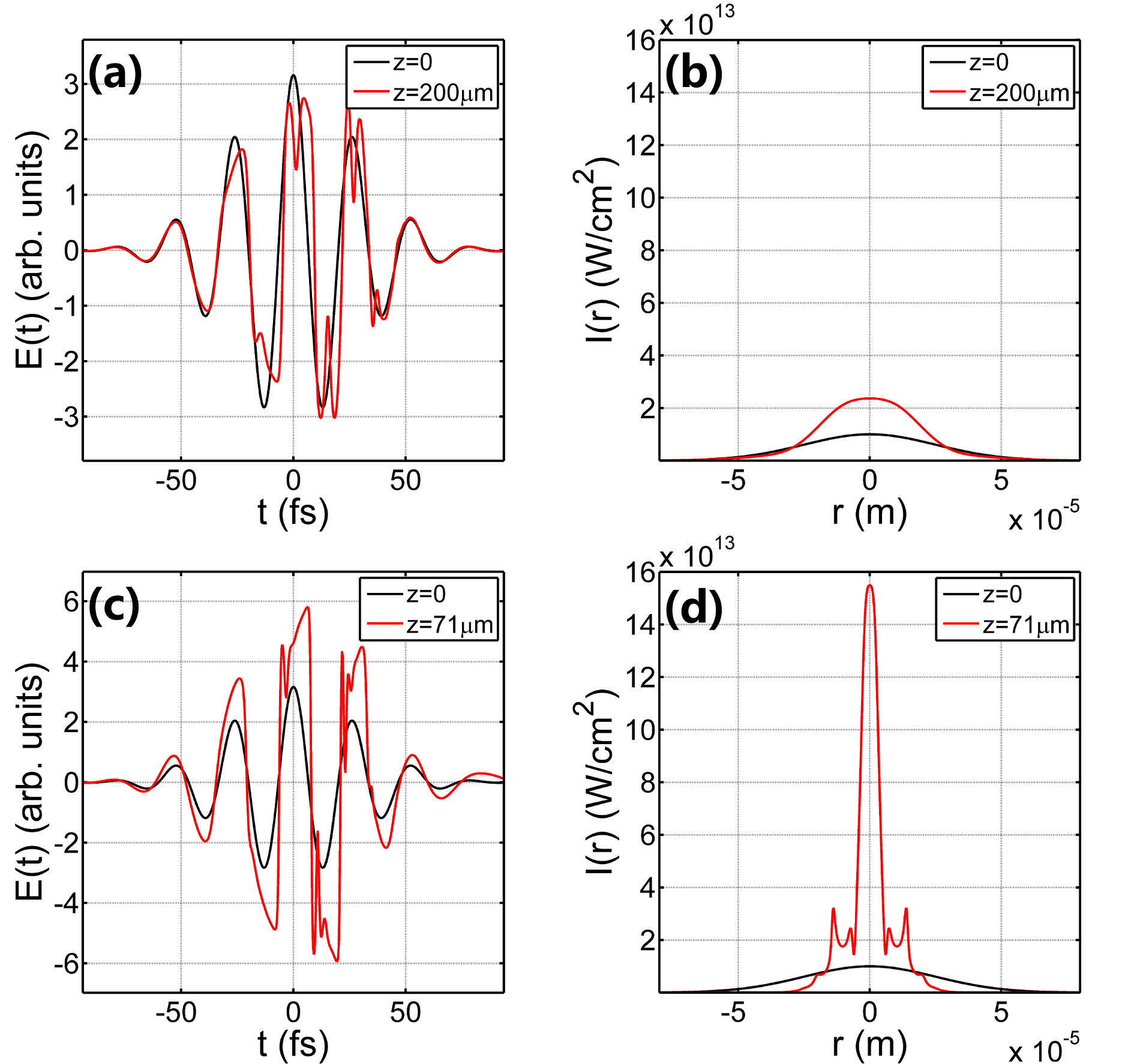}
\caption{(Color online) Electric field (first column) and beam waist (second column) for collimated (first line) and focused (second line) beam in single-crystal diamond. Black line: initial condition, red line: after propagation.}
\label{Fig_6}
\end{figure}

As shown in our previous work \cite{Whalen2014}, focusing conditions can play an important role in the carrier wave shock formation and evolution. In Fig. \ref{Fig_6} we can see the effect of an initial wavefront curvature on shock formation of our wavepacket propagating in single-crystal diamond, both in the temporal domain (a, c) and the spatial domain (b, d). Focusing the beam with a curvature of $f=50$ $\mu$m (c, d) will lead to earlier carrier shock formation and a top-hat field at $z=71$ $\mu$m. In contrast, the collimated beam is still evolving at $z=200$ $\mu$m, and has not reshaped into a top-hat yet. The obvious explanation of this behavior is the faster intensity increase due to strong linear focusing, which accelerates the harmonic generation and field steepening. Similar control over the field shape was predicted for Xenon in \cite{Whalen2014}, and is also observed in air and the rest of the noble gases (data not shown). Note that Fig. \ref{Fig_6} was generated using the UPPE model with ionization switched off. This was done to eliminate changes in the beam profile due to nonlinear losses and plasma defocusing. 

The transition from a collimated beam to a focused one has further implications in the spatial domain as well. As we can see in Fig.~\ref{Fig_6} (d), the initial curvature leads to strong spatio-temporal coupling. The formation of a ring around the on-axis peak, a feature not observed in the collimated case, is shown in Fig.~\ref{Fig_6} (b) and indicates steepening and breakup of the electric field.

\begin{figure}
\includegraphics[width=8.6 cm]{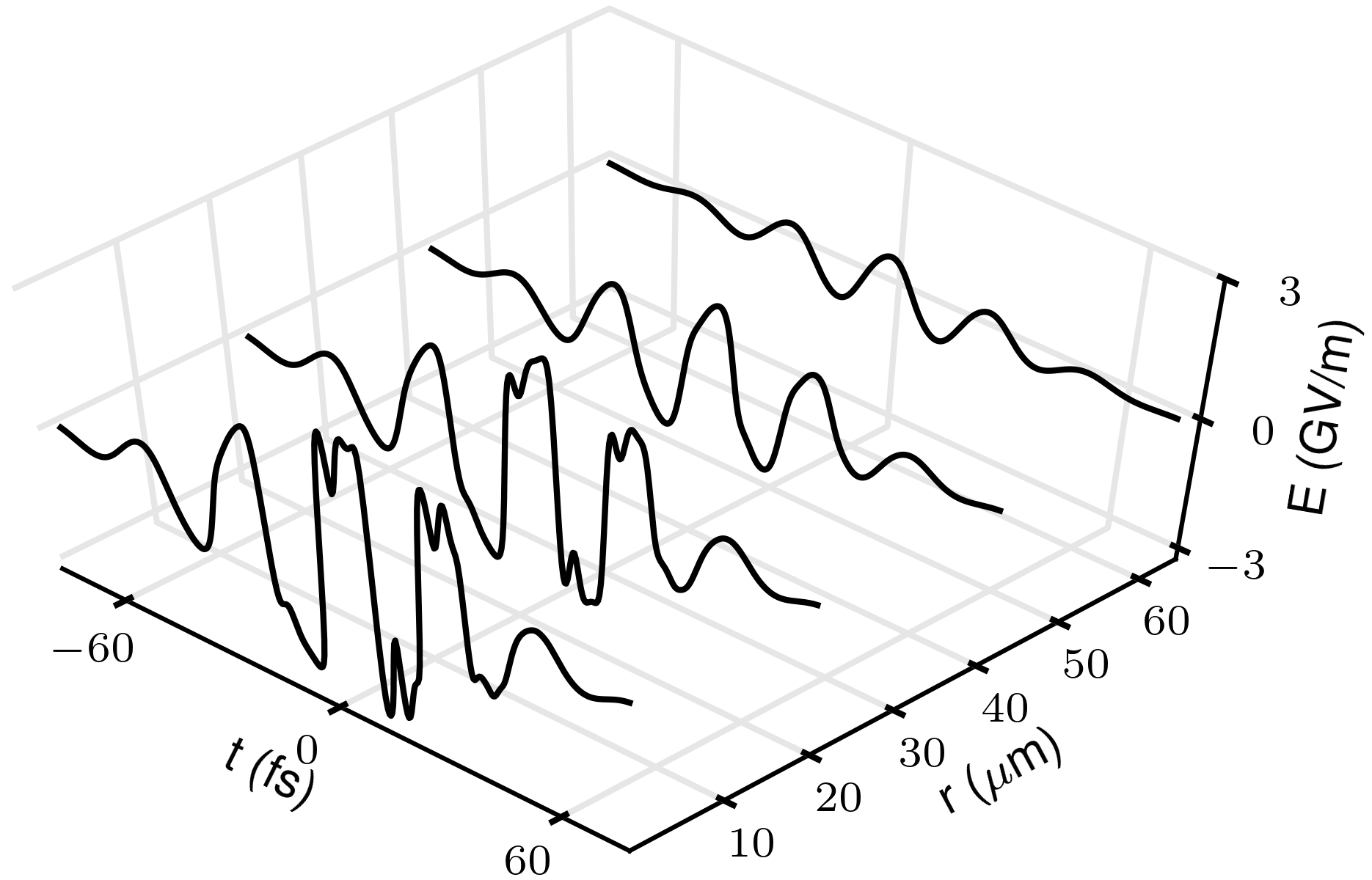}
\caption{Electric field at several radial positions after propagating a $\lambda_0 = 8$ $\mu$m collimated Gaussian beam for 170 $\mu$m in single-crystal diamond. }
\label{Fig_evolveradial}
\end{figure}

Note that up to now, all temporal field profiles are plotted on-axis, however at different positions in the radial dimension we expect to have more or less shock formation in the temporal electric field profile. Fig. \ref{Fig_evolveradial} shows the electric field of a wavepacket, propagated 170 $\mu$m in diamond without any initial curvature, as a function of the radial coordinate. As we can clearly see, the amount of field steepening is directly proportional to the field amplitude, and therefore varies in the radial dimension. The center part of the beam is showing the "top-hat" field shape, while as we move outwards the field steepening progressively weakens. At the outer part of the beam, the field retains it sinusoidal shape almost unaltered. This trend should be taken into account for cases where the beam profile undergoes spatial reshaping, as in the case shown in Fig. \ref{Fig_6} (b, d).

\section{Field shock characterization}
\label{sec:fieldShockChar}

In this section a detailed study and characterization of the carrier wave shock will be conducted.
\begin{figure}
\includegraphics[width=8.6 cm]{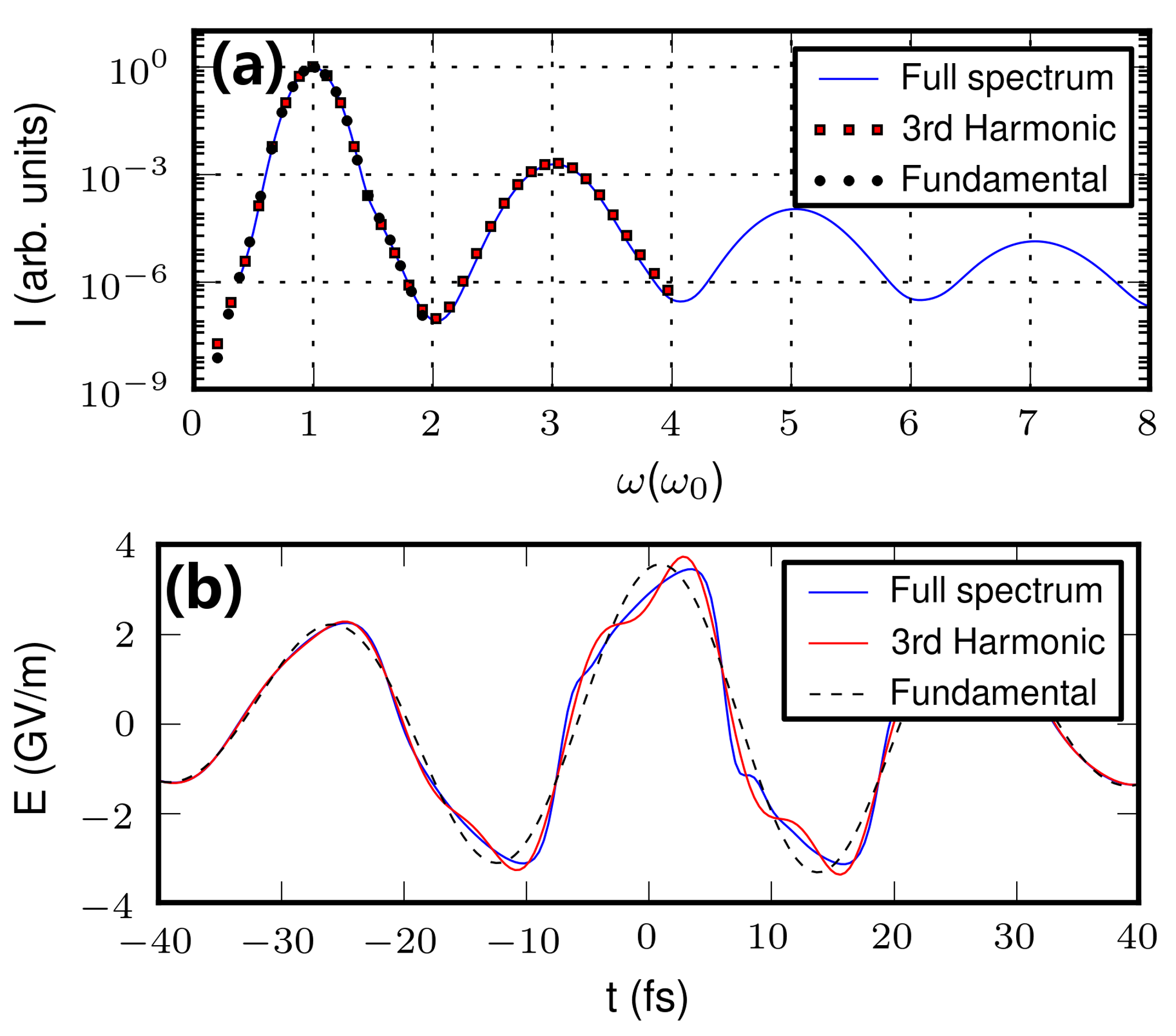}
\caption{(Color online) (a) Broadened spectrum of the 8-$\mu$m wavepacket after propagation in single-crystal diamond (cyan continuous line). Dashed lines: artificially truncated spectra, and (b) corresponding electric fields.}
\label{Fig_7}
\end{figure}
As was stressed in our previous work \cite{Whalen2014}, the main contributor to the formation of a carrier shock is the generation and co-propagation of the third harmonic wave. Here we will analyze and justify this claim in more detail. The carrier shock dispersion length scale $L_{CSD}$ was introduced in \ref{sec:mkp} as a natural length scale of the MKP equation. We also demonstated that $L_{CSD}$ is directly related to the third harmonic coherence length. In Fig.~\ref{Fig_7} (a) we can see the generated higher harmonics of the 8-$\mu$m beam in diamond, which is represented by the cyan continuous line. The corresponding field shape is shown in Fig.~\ref{Fig_7} (b) again with the cyan continuous line. To isolate the effect each of the harmonics has on field shape, we artificially truncated the spectrum up to the 3rd harmonic, shown in Fig.~\ref{Fig_7} (a) with the red squares. The corresponding field shape, shown in Fig.~\ref{Fig_7} (b) (red continuous line), is much more jagged when compared to the case where all the harmonics are taken into account. However, the electric field is still steepening, and the shock is still observable. In addition, the field shape is only slightly changing as the number of generated higher harmonics is varied from 1 to the full spectrum. On the other hand, if we keep only the fundamental by truncating all higher harmonics (including the 3rd), the field shape remains virtually unmodified as shown in Fig.~\ref{Fig_7} (b) (black dashed line). This clearly demonstrates that the actual steepening of the field is mainly coming from the third harmonic, while the rest of the harmonics smooth out the jagged oscillations observed when only the 3rd harmonic is taken into account. The oscillations on the field profile are interpreted as temporal interference fringes between multiple harmonics. In the case of the full spectrum, the large number of harmonics decrease the period of the fringes, effectively smoothing out the field profile. 

The above observation validates our claim that the carrier shock dispersion length $L_{CSD}$, as it was introduced first in \cite{Whalen2014} and derived in detail in \ref{sec:mkp}, is in fact a very robust way to predict shock formation by taking into account the material dispersion and laser pulse parameters. $L_{CSD}$ is a way to measure the length over which the carrier field shock dissipates due to dispersive effects of the medium. The main characteristic lengths, besides $L_{CSD}$, that are of interest here are the nonlinear length scale $L_{NL}$ and collapse distance $L_{C}$ as defined in Ref. \cite{Marburger1975},
\begin{equation}
L_{C}=\frac{0.367 L_{DF}}{\sqrt{((P_{in}/P_{cr})^{1/2}-0.852)^2-0.0291}},
\label{Eq_LC}
\end{equation}
\noindent where $P_{in}$ is the total input power and $P_{cr}$ is the critical power for self-focusing depending on the shape of the wave packet. For Gaussian spatio-temporal wave packets it is approximately given by $P_{cr}=3.77 \lambda_{0}^2/(8 \pi n_0 n_2)$. For all materials in this study, both $L_{NL}$ and $L_{C}$ are related to the instantaneous optical Kerr effect.

\begin{figure}
\includegraphics[width=8.6 cm]{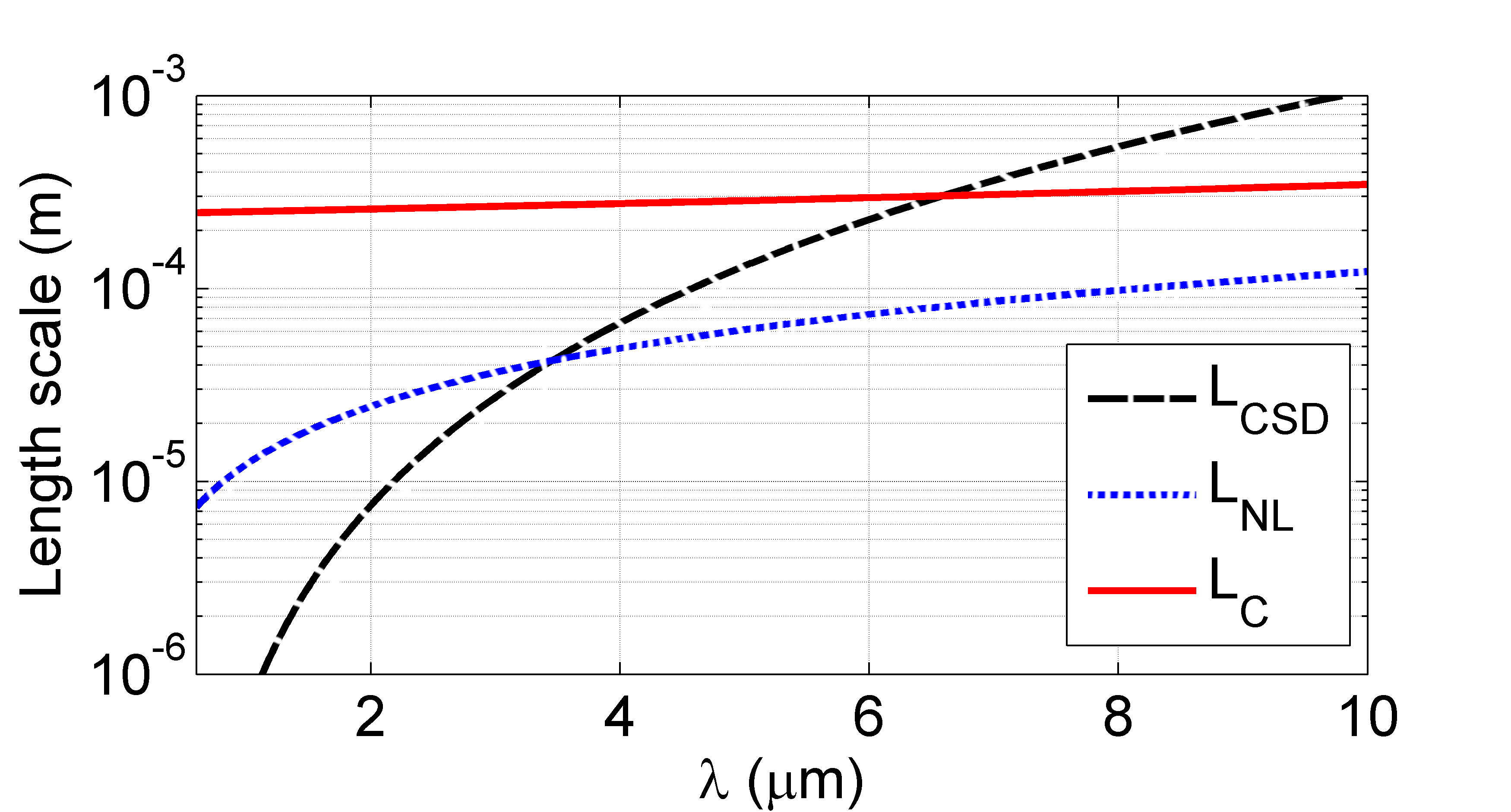}
\caption{(Color online) Characteristic length scales for the nonlinear Kerr effect (blue dotted line), carrier shock dispersion (black dashed line), and collapse distance (red continuous line) in single-crystal diamond as a function of wavelength. The input intensity is $1\times 10^{13}$ W/cm$^2$.}
\label{Fig_8}
\end{figure}

It is now possible to directly compare the significance of each physical effect for a given intensity value by their characteristic lengths. In Fig.~\ref{Fig_8} we can see all three characteristic lengths plotted for a peak intensity of $I_0=1 \times 10^{13}$ W/cm$^2$ (a typical value in the filamentation regime) as a function of wavelength in single-crystal diamond. The propagation of the wave packet is governed by the physical effect with the shortest characteristic length. The general trend is that at shorter wavelengths carrier shock dispersion is strong and the generated harmonics walk-off before field steepening can occur. At longer wavelengths, where dispersion is much weaker, the walk-off of the higher harmonics is slow, and a carrier wave shock is expected to occur. 

\begin{figure}
\includegraphics[width=8.6 cm]{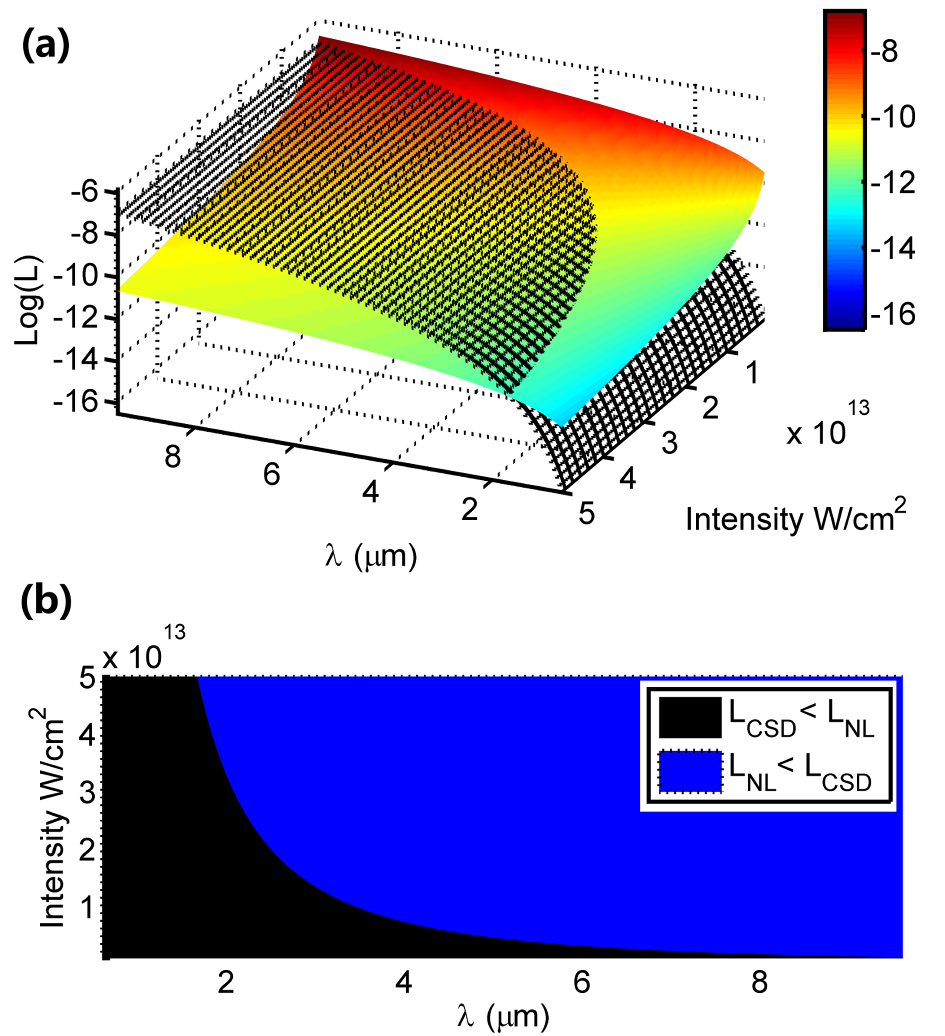}
\caption{(Color online) (a) Characteristic length surfaces of single-crystal diamond plotted in 3D. Black grid: $L_{CSD}$, color-coded surface: $L_{NL}$. (b) Minimum length scale over the $I-\lambda$ plane. Black region: $L_{CSD}<L_{NL}$ and blue region: $L_{NL}<L_{CSD}$ indicative of shock formation.}
\label{Fig_8b}
\end{figure}

Note that the carrier shock dispersion length scale is independent of the intensity. On the other hand, $L_{NL}$ depends on intensity, which for practical reasons makes a direct comparison between the two difficult due to the ever shifting peak intensity found in the filamentation regime. For this reason it is helpful to scan over a range of intensity values and produce a surface for each $L_{NL}$, and $L_{CSD}$, the latter being invariant along the intensity-axis. Fig.~\ref{Fig_8b}(a) shows the overlapping surfaces for $L_{NL}$ (color coded surface) and $L_{CSD}$ (black grid) for single-crystal diamond. Since the shortest length scale dominates, carrier shock is expected to manifest itself in areas where the surface corresponding to $L_{NL}$ is under the surface corresponding to $L_{CSD}$. To simplify the 3D surfaces shown in Fig.~\ref{Fig_8b}(a) we project them onto the $I-\lambda$ plane. In Fig.~\ref{Fig_8b}(b) we can see that the $I-\lambda$ plane is now separated in two regions, a region where carrier shock is expected to occur (blue) and a region where it is not (black), depending on which of the relevant length scales is shorter. The boundary between the two regions is given by the relation
\begin{figure}
\includegraphics[width=8.6 cm]{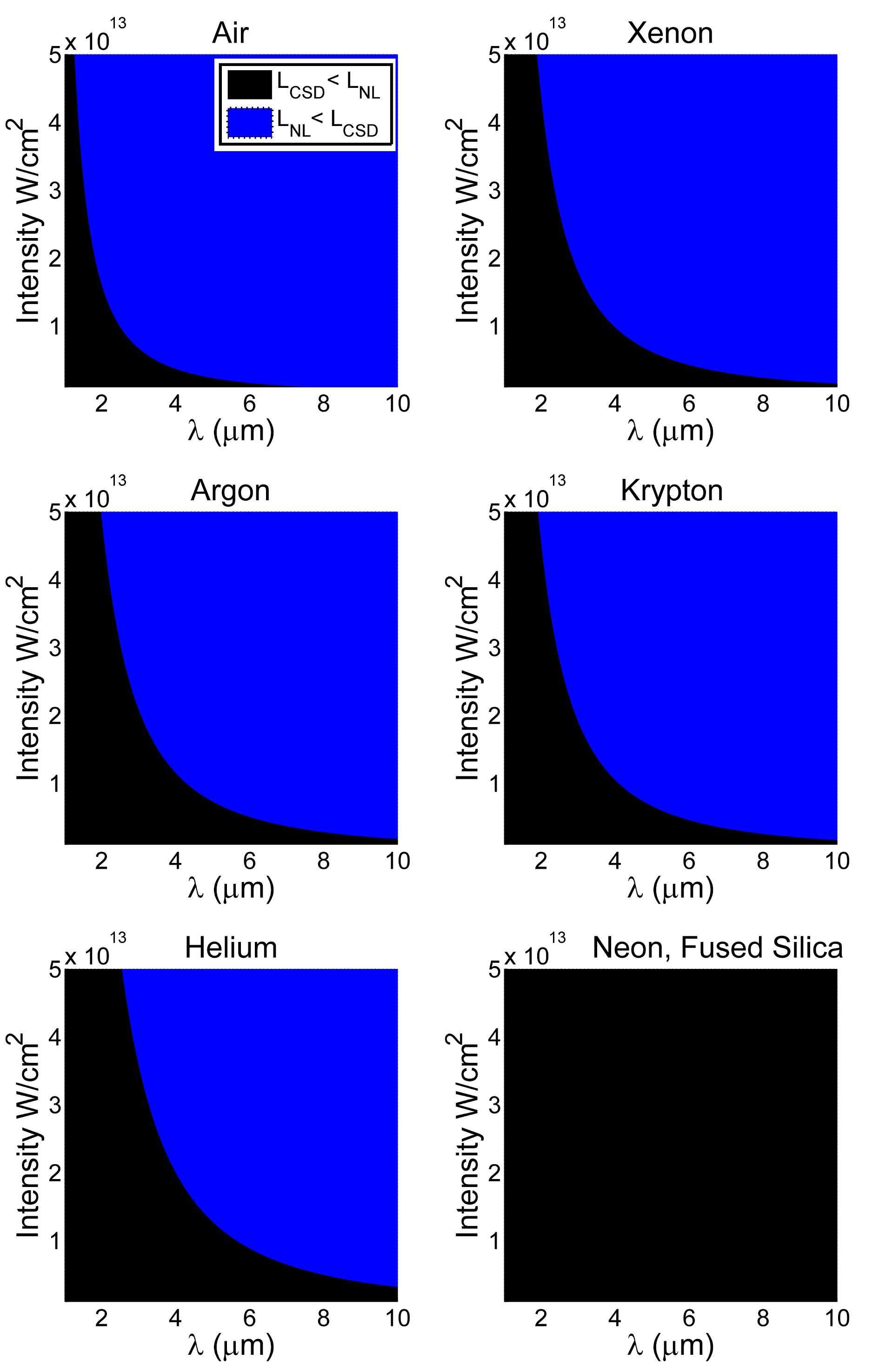}
\caption{(Color online) Minimum length scale over the $I-\lambda$ plane for air, xenon, argon, krypton, helium, neon and fused silica. Black regions: $L_{CSD}<L_{NL}$ and blue regions: $L_{NL}<L_{CSD}$ indicative of shock formation. }
\label{Fig_materials}
\end{figure}
\begin{equation}
I(\omega)=\frac{c \omega |k^{''}(3\omega)-3 k^{''}(\omega)|}{n_2}
\end{equation}
\noindent when $L_{CSD}=L_{NL}$. Fig.\ref{Fig_8b} provides a practical estimate for whether or not carrier shock formation is expected to happen in diamond.

\begin{figure}
\includegraphics[width=8.6 cm]{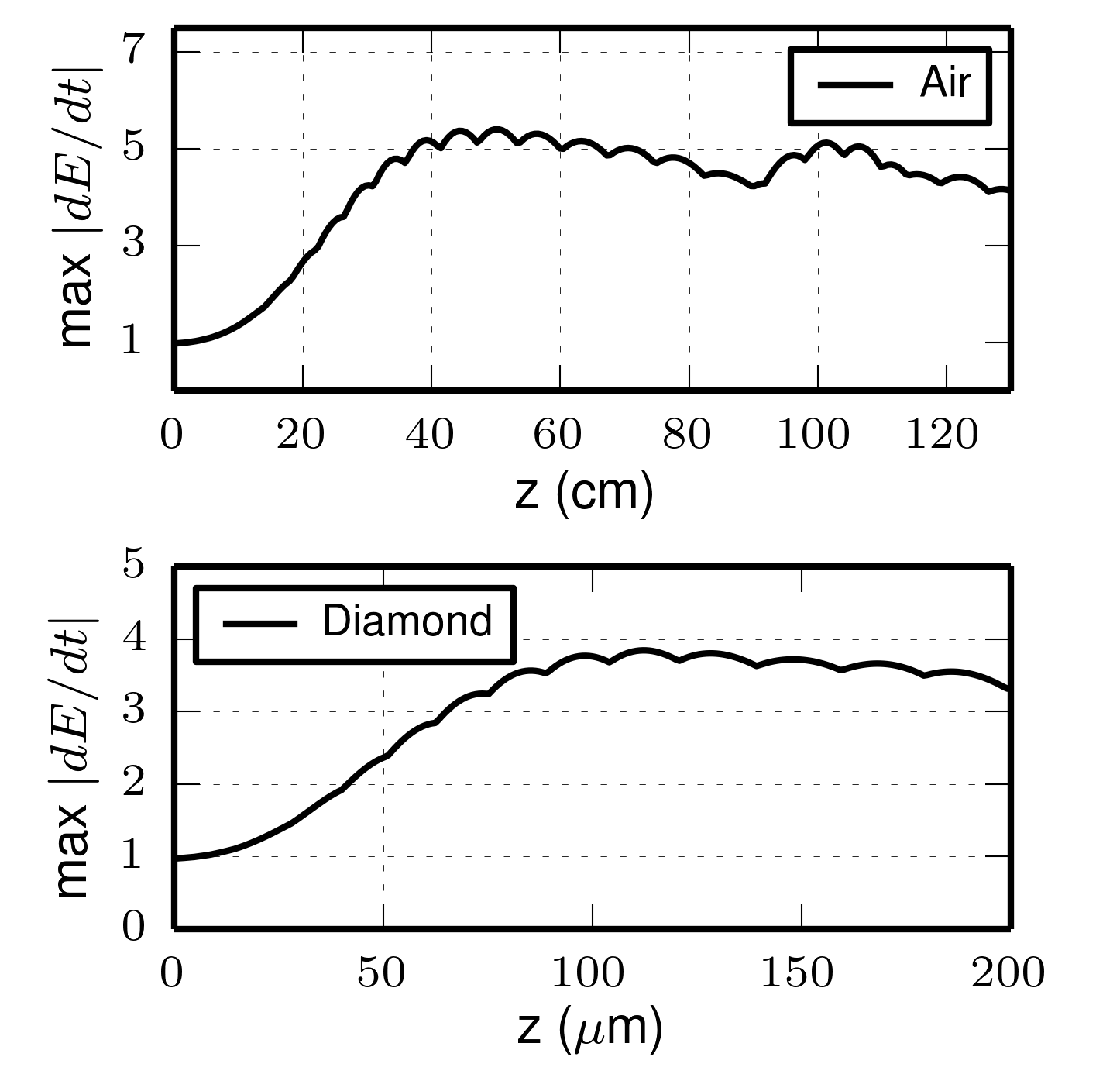}
\caption{Maximum of derivatives of the electric field for air (top) and single-crystal diamond (bottom). Propagation is simulated independent of a radial coordinate such that self-focusing and diffraction do not impact the results. }
\label{Fig_9}
\end{figure}

The comparison between $L_{CSD}$ and $L_{NL}$ shown in Fig.~\ref{Fig_8b} (b) can be done for any material with known dispersive and nonlinear coefficients. Fig.~\ref{Fig_materials} shows the $I-\lambda$ planes, separated into the "shock region" ($L_{NL}<L_{CSD}$) in blue and the "collapse region" ($L_{CSD}<L_{NL}$) in black, for all noble gases, air, and fused silica glass. As we can see, each material has a distinct region where carrier wave shock is expected to be observed. Furthermore, we emphasize that these regions of carrier wave shock are entirely determined by the dispersive and nonlinear properties of the medium. Note that not all materials are suited for field shock generation, both neon and fused silica have $L_{CSD} < L_{NL}$ in the $I-\lambda$ region of interest here. Neon and fused silica represent two extremes, the former being very weakly nonlinear, the later being very dispersive. It should be stressed that $L_{CSD}$ is a length scale corresponding to the MKP equation in a normal dispersion regime and that it primarily quantifies how flat the dispersive landscape is at a given carrier wavelength. For media with anamalous dispesion, such as fused silica, there are wavelengths where $L_{CSD}$ is no longer a useful predictor of shock \cite{Gilles1999}. 

Since carrier wave shock formation is mainly driven by competition between nonlinearity and dispersion, the latter being a fixed material property, it is expected that the amount of steepening in the electric field will vary along the propagation length, following the peak intensity to some extent. In Fig.~\ref{Fig_9} we can see the maximum of the temporal derivatives of the electric field as a function of propagation distance for air and single-crystal diamond. 

The slight field steepness saturation after propagating 50 cm in air, and 120 $\mu$m in diamond, occurs because the harmonics constantly walk-off. The field undergoes self-steeping a multitude of times as is indicated by the numerous oscillations for both materials. This happens because of the dynamic interaction between harmonic generation and walk-off. The oscillation period varies with intensity and material dispersion but is always observed. Similar results have been seen in all materials studied here (data not shown) as well an in the preceding work \cite{Whalen2014}, indicating that the phenomenon quite robust.

\section{Robustness and universality}

\begin{figure}
\includegraphics[width=8.6 cm]{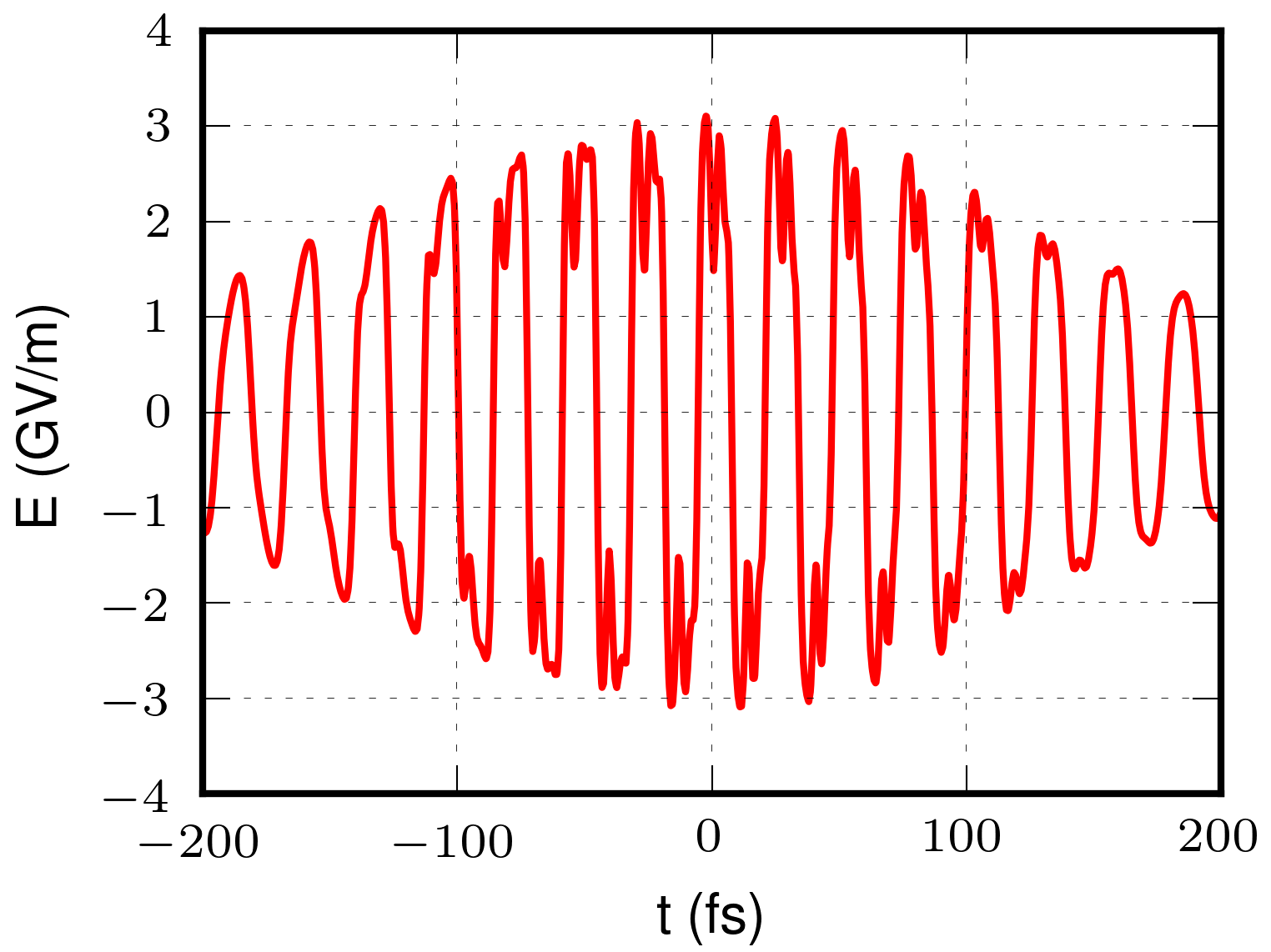}
\caption{(Color online) Electric field of a $\tau_p = 200$ fs, $\lambda_0 = 8$ $\mu$m pulse, propagated in single-crystal diamond for 170 $\mu$m.}
\label{Fig_10}
\end{figure}

A very important observation is that the carrier shock formation is unaffected by the pulse duration since neither $L_{CSD}$ nor $L_{NL}$ depend on it. In Fig.~\ref{Fig_10} we can see the electric field of a 200 fs pulse undergoing self-steepening in diamond after 170 $\mu$m of propagation. The carrier wave shock forms over multiple oscillations of the electric field. Interestingly, the reshaping of the electric field is not constant over the whole pulse, but rather the steepening follows the field amplitude. Oscillations at the front and back of the pulse show very little steepening, while as one moves closer to the central time slice, "shark-fin" like features become evident. At the center of the pulse the electric field is strongly reshaped into a "top-hat", effectively having gone through all stages of the carrier shock formation. As stated before, the carrier shock formation is intensity (or else field amplitude) related and is acting on each point of the field separately through the co-propagation of the generated harmonics. Therefore, the overall pulse duration is of little importance concerning carrier shock formation. However, the use of longer duration high power pulses will be limited by optical damage in solid state media like single-crystal diamond.

\begin{figure}
\includegraphics[width=8.6 cm]{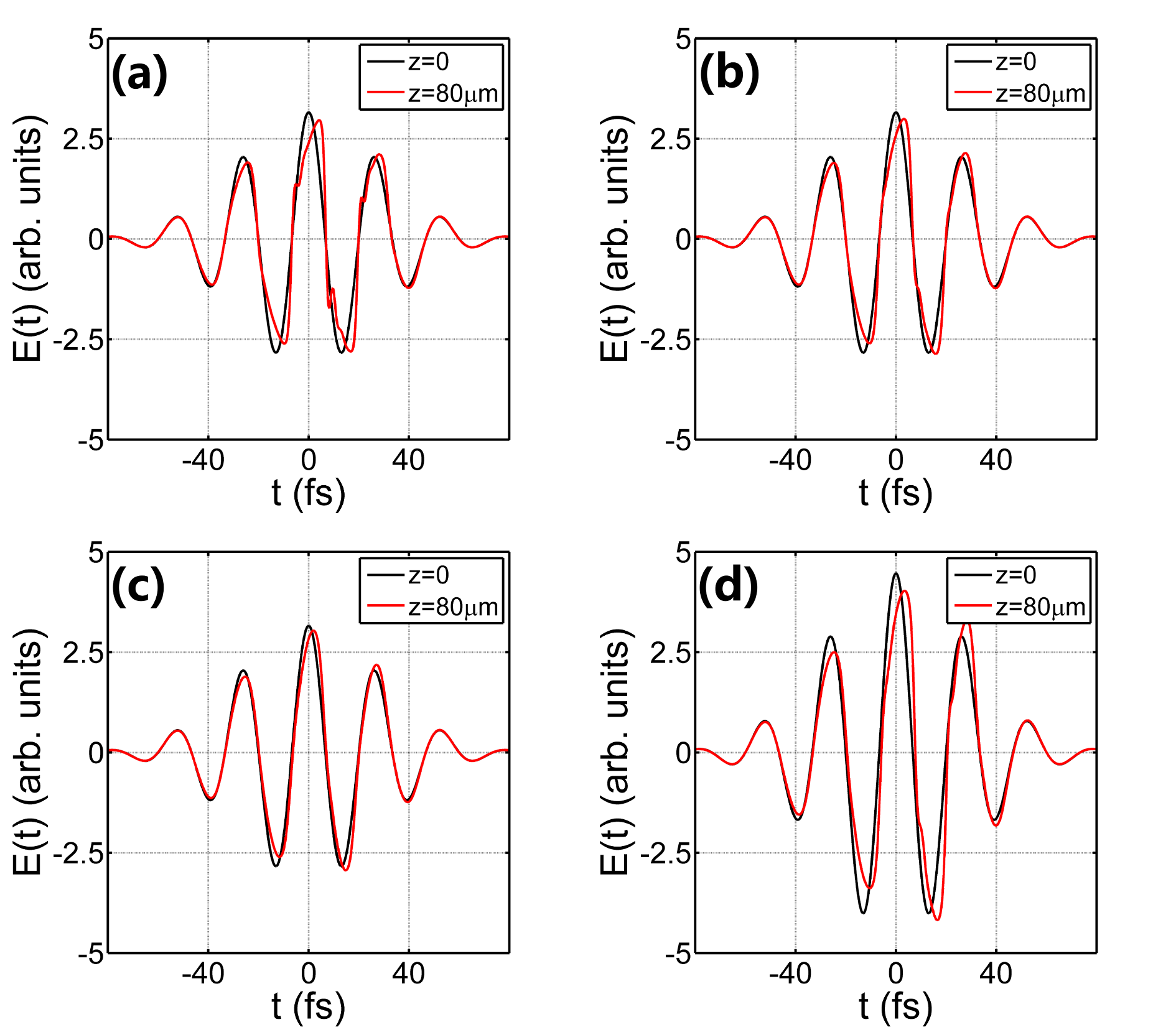}
\caption{(Color online) Electric field profiles for various Raman strengths after propagation of $z=80$ $\mu$m in single-crystal diamond. Raman contribution of (a) 0\% , (b) 25\%, (c) 50\%, and (d) 50\% with double the intensity value used in (a)-(c).}
\label{Fig_11}
\end{figure}

As mentioned earlier, we will investigate the role of vibrational Raman in the case of single-crystal diamond using the model described in section III. In order to isolate the effect of vibrational Raman optical field ionization, plasma is switched off. Simulations were conducted using the UPPE model. Conclusions drawn here should apply to all materials with a clearly defined Raman response. In Fig.~\ref{Fig_11} we can see the effect of vibrational Raman on the carrier shock formation of an $8$ $\mu$m - 40 fs pulse in single-crystal diamond. Although the frequency and time scale are known \cite{DiaRaman}, the actual relative strength (in comparison to the instantaneous Kerr effect) at the given wavelength is not. Therefore the relative strength is varied within a reasonable range from 0\% (Fig.~\ref{Fig_11}(a)), to 25\% (Fig.~\ref{Fig_11}(b)), up to 50\% (Fig.~\ref{Fig_11}(c)). In Figs.~\ref{Fig_11}(a)-(c) we observe that as the Raman strength is increased the amount of carrier shock is decreasing. However, the effect of Raman on the field shock formation is mostly neutral rather than destructive. This becomes evident when we increase the intensity by a factor of two in the case of 50\% Raman strength, which effectively reverses the loss of field shock due to the inclusion of Raman. Therefore, simply increasing the intensity of the pulse should in principle still lead to significant shock formation, given that intensity is kept under the damage threshold of the medium. This would mean that given precise knowledge of the Raman relative strength, a correction has to be applied to the shock formation region in the I-$\lambda$ plane in Fig.~\ref{Fig_8b} and Fig.~\ref{Fig_materials}. This would result in a shift of the collapse region (black) upwards, effectively shrinking the shock-formation region (blue) by an amount depending on the strength of the Raman response. 

%Concerning carrier wave shock formation, Fig.~\ref{Fig_11} makes it clear that vibrational Raman can in principle be overcome by a simple increase in intensity. This makes the phenomenon very robust since field shocks are expected to be observable regardless of the existence of vibrational Raman in the material. The drawback to this is that higher intensities are required, which could lead to optical damage and the generation of dense plasmas. 

\begin{figure}
\includegraphics[width=8.6 cm]{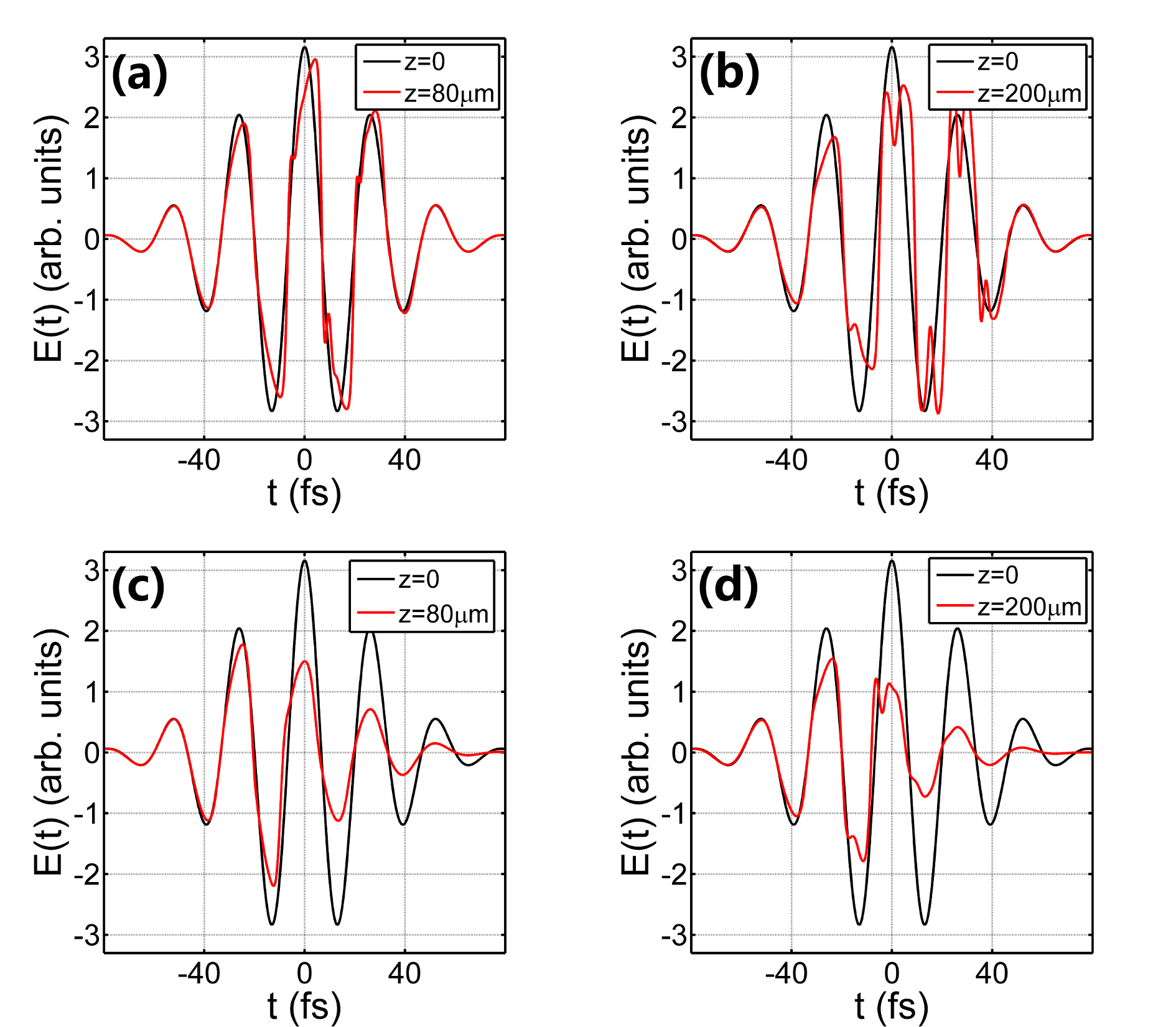}
\caption{(Color online) Electric field profiles of an $8$ $\mu$m - 40 fs pulse in single-crystal diamond, without (first row) and with ionization (second row). Black line: initial field shape, red lines: reshaped electric field after $z = 80$ $\mu$m (first column) and $z = 200$ $\mu$m of propagation (second column).}
\label{Fig_13}
\end{figure}

Lastly, we will study the effect of high nonlinear losses and plasma, including plasma defocusing and absorption, on carrier shock formation. Even though ionization is properly modeled as shown in section IIB. and Fig.~\ref{Fig:keldysh}, an accurate and well established ionization model for single-crystal diamond at mid-IR wavelengths is not available in the literature. Therefore, we investigated and compared carrier wave shock formation for two extreme cases: using artificially high ionization coefficients and switching ionization off altogether.

In Fig.~\ref{Fig_13} we observe the effect of artificially increasing ionization on the carrier shock formation of our 8$\mu$m - 40 fs wavepacket in single-crystal diamond. As we can see, the "shark-fin" and "top-hat" shaped electric fields are significantly reduced in amplitude (Fig.~\ref{Fig_13} (c), (d)) when compared to the case without ionization shown in Fig.~\ref{Fig_13} (a), (b). However, even with artificially high ionization coefficients, field steepening is still observed. Ionization losses tend to reduce field amplitudes, and consequently they reduce the amount of field steepening. On the other hand, plasma generation has been associated with a strong blue-shift in the spectrum and supercontinuum generation \cite{Couairon1}, which could in principle contribute to field shock formation in a similar way as higher harmonic generation does. The effect of plasma on field shock formation is proving to be a complex process, depending on setup parameters and propagation dynamics. We have to stress again that the use of shorter wavelengths will lead to an increase of ionization, which along with stronger dispersion makes carrier shock formation extremely difficult to observe. On the opposite side, moving towards longer wavelengths favors field shock formation, since ionization is weaker and $L_{CSD}$ becomes longer.

\section{Comparison between GMKP and UPPE}

\begin{figure}
\includegraphics[width=8.6 cm]{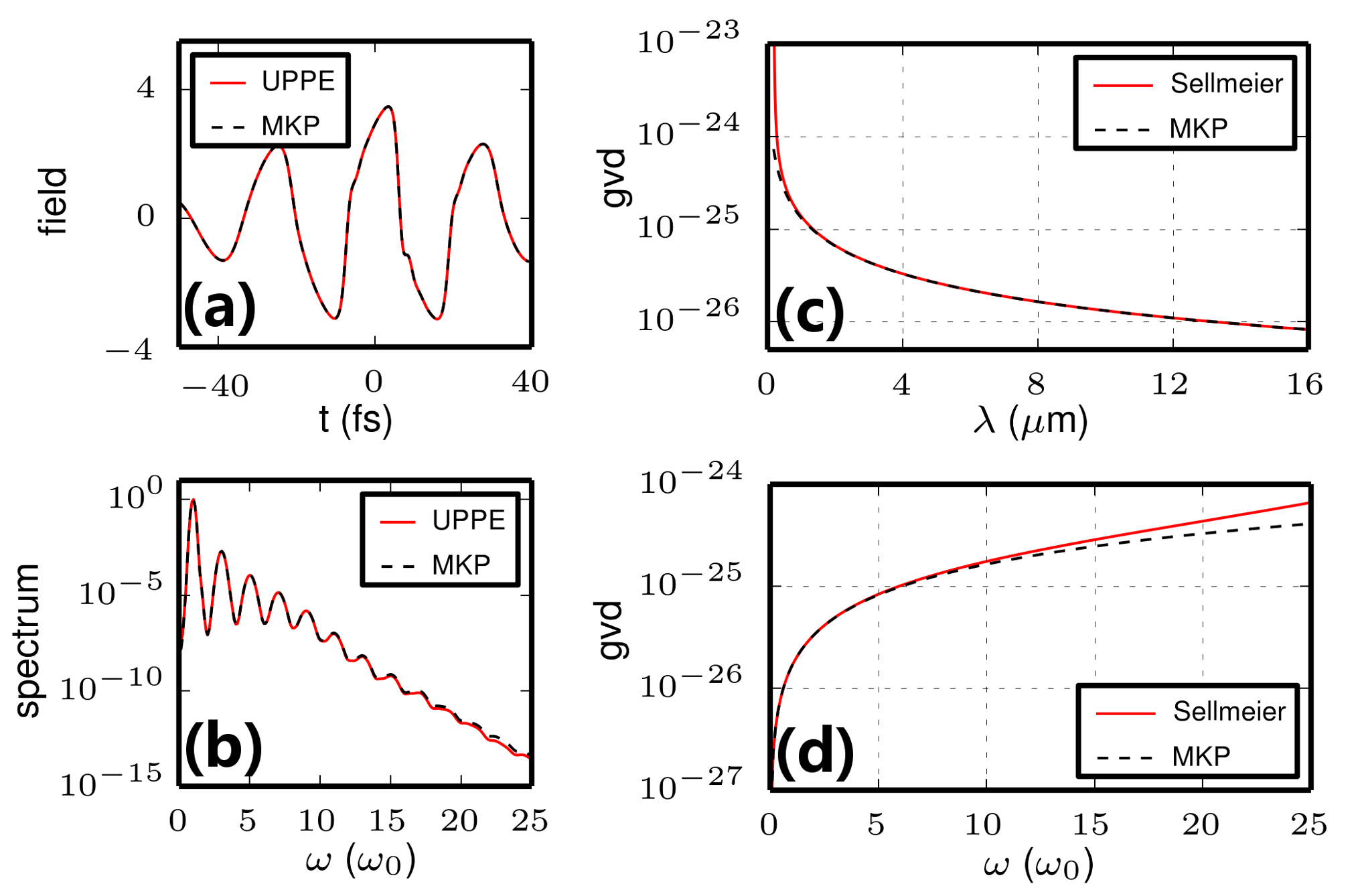}
\caption{(Color online) Comparison of carrier shock formation between GMKP (black dashed lines) and UPPE (red continuous lines), for an 8 $\mu$m - 40 fs wavepacket in single-crystal diamond. (a) Generated "shark-fin" electric field for UPPE and GMKP. (b) Corresponding spectral intensities. (c) Group velocity dispersion as modeled by GMKP and UPPE for single-crystal diamond vs wavelength $\lambda$, and (d) vs normalized frequency.}
\label{Fig_14}
\end{figure}

To validate our result we compared the results of the GMKP model with the full UPPE field propagator throughout this work. In Fig.~\ref{Fig_14} we can see minor discrepancies between the two models in the case of our 8 $\mu$m - 40 fs wavepacket propagating in single-crystal diamond. For simplicity, no vibrational Raman or optical field ionization were taken into account. In Fig.~\ref{Fig_14} (a) we can see the predicted carrier wave shock using both the GMKP (blue lines) and UPPE (black dashed lines) models. The corresponding spectral intensities are shown in Fig.~\ref{Fig_14} (b). The field profiles are essentially indistinguishable, showing that at long wavelengths the use of the GMKP model is fully justified. In Figs.~\ref{Fig_14} (c) and (d), the dispersion relation of diamond is shown for both models, visualized in wavelength and normalized frequency respectively. As we can see there is a slight discrepancy between the GMKP dispersion and the actual Sellmeier dispersion relation used in the UPPE model, especially at high frequencies. This should be taken into account when conducting numerical experiments at shorter wavelengths. For shorter wavelengths, the GMKP model can poorly approximate material dispersion and hence significant differences between the UPPE and GMKP models may result. This means that the pulse's wavelength can determine the numerical model of choice, UPPE for shorter wavelengths where dispersion is accurately modeled and GMKP at longer wavelengths since it is the canonical equation describing pulse propagation in that regime \cite{Whalen2014}.

\section{Conclusion}
We have presented a broad and detailed numerical study of carrier wave shock formation in the weakly dispersive - highly nonlinear regime in air, noble gases and single-crystal diamond. "Shark-fin" and "top-hat" electric fields are predicted to form spontaneously at long wavelengths for a variety of input powers and pulse durations. In addition, the field shock formation is proven to be very robust since it is occurring despite ionization losses, plasma generation, vibrational Raman and two-phonon absorption. Carrier shock is predicted to be unavoidable in most high intensity, long wavelength pulse propagation settings. The results of the GMKP model were compared with the ones from the full field UPPE model, and were found to be in good agreement with each other. Results presented here are expected to have significant impact on multiple disciplines in the field of nonlinear optics, especially attosecond pulse generation \cite{attorev}, waveform synthesis \cite{Science2011}, and pulse compression \cite{Berge2013}.

\section*{Acknowledgments}
This work was supported by an Air Force Office of Scientific Research Multidisciplinary University Research Initiative (MURI) grant  FA9550-10-1-0561.

\end{document}